# Effect of Austerity Measures on Infant Mortality: Evidence from Greece[*]

Robert John Kolesar          Rok Spruk

## Abstract

*This study examines the effect of fiscal austerity measures on infant mortality in Greece. Austerity measures were initiated by the tripartite committee and implemented between 2010 and 2017 to counteract the country's deep fiscal deficit and large public debt. By comparing Greece with a plausible donor pool of OECD and Union for the Mediterranean member states in the period 1991-2020, we estimate a series of missing counterfactual scenarios to evaluate the infant mortality effects of large-scale reduction in spending on health care. A series of hybrid synthetic control and difference-in-differences estimates indicate a unique and pervasive increase in infant mortality after the implementation of austerity measures. Compared to a plausible OECD and Mediterranean counterfactual scenario, pro-cyclical austerity measures are associated with derailed and permanently increased infant mortality up to the present day. Our estimates suggest that compared to a plausible counterfactual scenario, the cumulative infant mortality cost of austerity policies exceeds 10,000 infant deaths or slightly less than 850 deaths for each year of the austerity policies. Notably, mortality increases are concentrated among boys. The estimated impacts survive a battery of rigorous robustness and placebo tests.*

---

[*] Kolesar: Deputy Director for Policy and Financing, Paladium Group, 1331 Pennsylvania Avenue, NW Suite 600, Washington D.C., 20004, United States of America. Spruk (corresponding author): Associate Professor of Economics, School of Economics and Business, University of Ljubljana, Kardeljeva ploscad 17, SI-1000 Ljubljana. E: rok.spruk@ef.uni-lj.si.

# 1 Introduction

Achieving and maintaining vital public health outcomes and equitable service access necessitates the large-scale investment in health systems, capacity and preventive care (Acemoglu and Johnson 2007). In turn, large fiscal policy shocks to either government expenditure and taxes may have both short-term and long-term consequences for public health outcomes that affect both equity, quality and provision of care (Corsetti 2012; Bailey and Goodman-Bacon 2015; Fadlon and Nielsen 2021; Romer 2021). On the eve of the financial crisis in 2009, Greece enjoyed one of the most enviable levels of public health in the OECD reinforced by high life expectancy and one of the world's lowest rates of infant mortality. In 2009, Greece's infant mortality rate stood at around three deaths per 1,000 live births, or around one half of the average mortality rate of OECD members. In the same year, Greece experienced a severe and prolonged economic crisis. The Eurozone's public debt crisis began in Greece in 2009 (Chodorow-Reich, Karabarbounis, and Kekre 2023). To avoid bankruptcy, Greece had to borrow €256.6 billion from the International Monetary Fund (IMF), European Union, and the European Central Bank, under strict conditions that included sharp reductions in government spending (Kentikelenis et al. 2011; Gourinchas, Philippon, and Vayanos 2017). Greek public spending fell by 36 percent from 2009 to 2014 (Burki 2018; Commission, Economic, and Affairs 2023). The 2010 Troika Memorandum limited public expenditure on health to six (6) percent of Gross Domestic Product (Papademos, Venizelos, and Provopoulos 2012). By 2019, Greece was one of the few OECD member states having public health expenditure less than 4 percent of the GDP. Apart from Syria, Greece was the only nation in the Mediterranean basin and the only OECD member experiencing an absolute increase in infant mortality rate from 2009 until 2017.

The austerity measures introduced under the Troika framework included salary cuts and layoffs among public employees; reductions to minimum wages, pensions, and welfare payments; cuts to investments and public consumption; local administration reforms; increased value added tax (VAT); increase in specific fuel, tobacco, and alcohol taxes; special levies on corporate profits; increased taxes on real estate; and privatisation of state assets among others. These measures weakened the capacity of the system to address social risks such as unemployment, inequality and poverty while exacerbating structural problems in the healthcare system (Kentikelenis and Papanicolas 2012; Adam and Papatheodorou 2016; Charalampos Economou et al. 2015). There is evidence indicating that IMF programs have been significantly associated with "weakened health care systems, reduced effectiveness of health-focused development aid, and impeded efforts to control tobacco, infectious diseases, and child and maternal mortality" (Stuckler et al. 2017). The three successive Economic Adjustment Programmes implemented in Greece are estimated to have a long-term negative impact on the Greek GDP per capita and rates of GDP growth (Efthimiadis, Papaioannou, and Tsintzos 2013). Sharp declines in disposable income and the dramatic increase in unemployment led to a significant increase in absolute poverty (Mitrakos 2014). A systematic literature review found that high austerity is associated with increased material deprivation, child poverty rates, and low birth weight (Rajmil et al. 2020). There is also evidence that several expenditure-reducing reforms have long-lasting social consequences with serious detrimental effects on families with children (Kaplanoglou and Rapanos 2018).

In this paper, we examine the contribution of fiscal austerity to infant mortality and exploit the introduction of Economic Adjustment Programme in Greece to consistently estimate the contribution of austerity to public health outcomes. Our identification strategy relies on non-parallel trends in infant mortality dynamics as a valid source of the missing counterfactual scenario. By comparing Greece with a stable donor pool of OECD member states that have not undergone prolonged austerity policies amid the Economic Adjustment Program, we estimate the counterfactual scenario through the application of the synthetic control estimator (Abadie 2021). We show that the synthetic control estimator provides an excellent quality of the fit between infant mortality trajectories of Greece and its synthetic control group prior to the austerity. Our results show that relative to the synthetic counterfactual, troika-orchestrated public health care spending reductions are associated with a marked and permanently elevated trajectory of infant mortality of around 31 percent in the post-austerity period. Infant mortality increase is around one-third larger for boys and does not disappear in the years preceding the COVID-19 pandemic. The estimated increase in infant mortality survives an extensive battery of robustness checks and is robust against the composition of the donor pool, and does not appear to be driven by pre-existing trends, shocks and policies.

We address two distinctive voids in the literature. First, following the economic crisis, Greece underwent a notable increase in suicide-related deaths relative to the estimated counterfactual scenario (Kubrin, Bartos, and McCleary 2022). Our approach fills the void in the literature by showing that austerity measures have disproportionately strong impact on the most vulnerable demographic segments of the population. This is done by estimating a counterfactual benchmark for the evaluation of the infant and childhood mortality effects of sharp fiscal policy shocks. And second, our analysis disentangles whether the fiscal policy shocks produce either a permanent or temporary impact on public health outcomes. By disaggregating the effect by gender and extending the analysis to neonatal and post-neonatal mortality, we are able to provide additional evidence on gender-related and age-specific disparities in the public health effect of large fiscal policy shocks.

Several recent papers are highly related. First, Berman and Hovaland (2024) examine the contribution of austerity measures to excess mortality by exploiting the introduction of austerity policies in the United Kingdom in 2010. Using difference-in-differences based identification strategy, the authors uncover the evidence of 3 percent increase in overall mortality rate or the equivalent of 190,000 excess deaths between 2010 and 2019 alongside a notable slowdown in life expectancy that appears to be particularly strong for the adult male population (Berman and Hovland 2024). And second, Kubrin et. al. (2022) examine the relationship between austerity measures and suicide rates both for the population as a whole as well as for men and women. By utilizing a panel of 50 member states of World Health Organization, their analysis employs a synthetic control design to estimate the IMF-imposed austerity measures and suicide rates. The results show that austerity policies resulted in statistically significant suicide rate increase that has been particularly stronger for women relative to the synthetic counterfactual (Kubrin, Bartos, and McCleary 2022). Our study employs a similar quasi-experimental approach to identify the contribution of austerity policies to infant and child mortality rate and can be considered one of the first studies examining this link between austerity policies and mortality patterns among the most vulnerable segment of the population using the counterfactual approach.

This study shows that prior to the economic downturn and subsequent austerity measures of the Economic Adjustment Programme, Greece enjoyed both one of the lowest and continuously declining overall and gender-specific rates of infant mortality. Contrary to the expected slowdown of mortality reduction typically identified in the austerity period, this study shows that Greece experienced an absolute increase in the number of neonatal and post-neonatal deaths and the number of deaths before the age of five. Elevated mortality trajectories are exceedingly rare both in the OECD as a plausible peer group as well as in high-income countries more generally. Furthermore, this study permits the estimation of both the increase in the absolute number of deaths as well as the rate of mortality relative to the comparison group where such harsh austerity measures were not introduced. This allows us to properly estimate the hypothetical mortality trajectory driven by general downward trend and compare it with the observed trajectory, which allows us to quantify the mortality impact of the austerity measures.

We show that sharp reductions in public health spending during severe economic downturn immediately materialize into elevated mortality rates among infants and children as the most vulnerable demographic groups affected by the fiscal shock. Despite the temporary economic adjustment program put in place between 2010 and 2017, we find evidence of a protracted increase in infant mortality trajectories of both boys and girls relative to the synthetic control group. Notably, mortality increase are concentrated among boys. In spite of the some reduction in the rate of infant mortality after 2017, Greek trajectory has been elevated to such degree that it has not yet recovered to the levels predicted by its synthetic control . Against this backdrop, this study underscores the importance of continuous investment in public health systems and preventive care to preserve low rates of mortality as one of perhaps most fundamental pillars of enduring and sustainable human development. It also underscores the perils posed by sharp and large-scale fiscal shocks, demonstrating how the external orchestration of austerity measures materialized into deep expenditure cuts can easily undermine decades of progress in the equity, quality and access to public health services, particularly for the most vulnerable.

The rest of this article organized as follows. Section 2 reviews the literature and policy background. Section 3 discusses the data and sample. Section 3 presents the identification strategy. Section 4 presents and discusses the results and the corresponding robustness checks alongside with the discussion. Section 5 offers a few conclusive remarks.

## 2 Background

### 2.1 Austerity

Austerity measures refer to reduced social spending and increased taxation. Such measures have direct and indirect impacts on health. Direct austerity impacts include cuts to healthcare services, as well as reductions in health coverage and restricting access to care; these can be characterized as the 'healthcare effect' (Stuckler et al. 2017). For example, increased out-of-pocket expenditure can create an access barrier and undermine financial risk protection (Grigorakis et al. 2018). The World Health Organization found that increases in healthcare user fees and copayments led to nearly one-third of the poorest Greek households incurring catastrophic medical expenses in 2014 (Burki 2018).

Indirect austerity impacts relate to increasing unemployment, poverty, homelessness and other socio-economic risk factors; these can be characterized as the 'social risk effect' (Stuckler et al. 2017). For example, employment and improved working conditions can improve health and reduce gender-based health inequalities (Naik et al. 2019). In addition, higher unemployment is negatively related to individuals' health conditions – an increase in unemployment is associated with a significant increase in the probability of diabetes, infarction, ulcer, cirrhosis, and nervous disorders (Colombo, Rotondi, and Stanca 2018). Adult unemployment in Greece rose from 6.6 percent in May, 2008, to 16.6 percent in May, 2011 with youth unemployment rising from 18.6 to 40.1 percent over the same period. Concomitantly, increased morbidity and the inability to afford private healthcare led to increased utilization of public health services (Kentikelenis et al. 2011). This increased demand was complicated by reduced public expenditure (Kentikelenis and Papanicolas 2012). The interaction of fiscal austerity with economic shocks and weak social protection has been noted to escalate health and social crises in Europe (Karanikolos et al. 2013).

## 2.2 Health system effects

Prior to the 2009 economic crisis the Greek health system suffered from a wide range of challenges including fragmented systems and funding, reliance on informal payments, inequities in access, financing, supply and quality of services; the system was plagued by corruption, fraud, waste, patronage, and inefficiencies such as oversupply of specialists and undersupply of nurses (Mossialos, Allin, and Davaki 2005; Charalampos Economou et al. 2015; Charalambos Economou and Organization 2010; Davaki and Mossialos 2005; Commission, Economic, and Affairs 2023). Greek austerity measures included many national health system reforms aimed at improving efficiency and cost-effectiveness. These included the creation of the national health insurance fund (for resource pooling), harmonization of contribution rates, centralized procurement of supplies and other cost-savings measures such as base pricing, generic and electronic prescribing, and annual spending caps (Commission, Economic, and Affairs 2023).

The focus was on immediate expenditure reductions with uncertain long-term consequences for Greek public health and healthcare (Simou and Koutsogeorgou 2014). Draconian cuts to public healthcare expenditure had significant affects on the structure and functioning of public hospitals exasperating insufficient staffing, operating deficits, drug shortage and lack of basic medical supplies (Ifanti et al. 2013). For example, in 2010 health-care staff salaries were reduced twice; maternal and child health services were cut by 73 percent between 2009 and 2012; and, public hospitals funding was slashed by 50 percent from 2009 to 2015 (Burki 2018). Budget cuts were compounded by increased admission rates as people shifted from private providers and the uninsured population swelled (Kentikelenis et al. 2011; Adam and Papatheodorou 2016). Mortality due to adverse events during medical treatment is estimated to account for 242 excess deaths per month starting in September 2008 (Laliotis, Ioannidis, and Stavropoulou 2016).

The crisis also led to declining social insurance revenues as unemployment increased, wages fell, and working hours were cut (Burki 2018). This erosion of Greece's public health system is reflected in increased patient dissatisfaction with hospital care as well as unmet need for healthcare increasing from 10.0 to 21.9 percent following the adoption of austerity measures (Keramidou and Triantafyllopoulos 2018; Zavras et al. 2016; Carr and Wolfe 2019; Filippidis et al. 2017). A 2017 review of the Greek health systems found that the Economic Adjustment Programme acted as a catalyst for health reforms aimed at reducing public sector spending while reducing inequities and inefficiencies. Multiple reforms have been implemented since 2010. These include the establishment of a single purchaser, standardizing the benefits package, re-establishing universal coverage and access to health care, demand and supply-side measures to reduce pharmaceutical expenditure, and important changes to procurement and hospital payment systems (Charalampos Economou et al. 2017).

*2.3 Healthcare effect*

High income inequality is strongly associated with increased infant mortality (Siddiqi, Jones, and Erwin 2015). Avendano asserts that this association is not causal and suggests that "social policies that reduce infant mortality cluster in countries with low-income inequality, but their effects do not operate via income"; this highlights the need to examine the impact of specific social policies on infant mortality (Avendano 2012). For example, there is evidence that infant mortality is strongly associated with relative and absolute income as well as the unemployment rate (Dallolio et al. 2012). This is also evidence that stalled mortality trends in many high income countries is likely to be at least partly due to austerity policies (McCartney 2022).

There is important evidence demonstrating that the deterioration of self-reported health as well as health service access and availability in Greece has led to increased morbidity and mortality (K.N. Fountoulakis and Theodorakis 2014; Vandoros et al. 2013). Many studies have investigated the mortality effects of the Greek economic crisis (Kotzamanis, Zafeiris, and Kostaki 2022; Zafeiris and Kostaki 2019; Kentikelenis et al. 2014b; Doetsch et al. 2023; Granados and Rodriguez 2015; Rajmil et al. 2018; Zilidis and Hadjichristodoulou 2020; Siahanidou et al. 2019; Filippidis et al. 2017; K. Fountoulakis et al. 2022; Michas et al. 2014; Kentikelenis et al. 2014a;

Laliotis, Ioannidis, and Stavropoulou 2016). Cause-specific mortality is important as more generalized metrics such as life expectancy and overall mortality can mask importance changes (Granados and Rodriguez 2015; K. Fountoulakis et al. 2022). Several studies have found a slowdown in Greek life expectancy after 2010, and increased infant mortality, suicide, circulatory, infectious and parasitic diseases alongside a drop in road traffic accident mortality (Kotzamanis, Zafeiris, and Kostaki 2022; Filippidis et al. 2017). A systematic review of 39 studies on the effects of the economic crisis on health and healthcare in Greece from 2009 to 2013 found post-crisis deterioration of public health with increasing rates of mental health, suicides, and epidemics, and deterioration of self-rated health (Simou and Koutsogeorgou 2014).

### 2.4 *Conflicting findings on infant mortality*

There is evidence showing that countries adopting stricter austerity policies experience slower improvements in mortality (McCartney 2022). At least 12 published papers analyze mortality in Greece following the economic crisis (see Appendix 1). These studies used data from the Hellas Health surveys, the Hellenic Statistical Authority, the World Bank, the Organization for Economic Cooperation and Development, Eurostat, and the International Monetary Fund. Three (3) studies found differences in mortality by-type (Filippidis et al. 2017; Laliotis, Ioannidis, and Stavropoulou 2016; Kotzamanis, Zafeiris, and Kostaki 2022). Six (6) studies found increases in infant mortality following the onset of the economic crisis (Filippidis et al. 2017; Kotzamanis, Zafeiris, and Kostaki 2022; Doetsch et al. 2023; Zilidis and Hadjichristodoulou 2020; Siahanidou et al. 2019; K. Fountoulakis et al. 2022). Two (2) studies found a temporary increase in infant mortality with recovery (Michas et al. 2014; K.N. Fountoulakis and Theodorakis 2014). One study found no evidence of a health crisis in Greece (Granados and Rodriguez 2015). This study analyzed population health and health services metrics from 2003-2007 with 2008-2012. However, the crisis started in late 2009, raising questions about the periods compared. Finally, one (1) study found infant mortality decreased from 2008 to 2010 and then rebounded to levels similar to 2005-2007 in the 2012-2015 period (Rajmil et al. 2018). Again, the periods analyzed do not align well with the crisis.

We were only able to identify one mortality study (focusing on suicide) that applies quasi-experimental methods (Kubrin, Bartos, and McCleary 2022). Most studies have focused on the short-term effects owning to the timing of the research as the consequences of a population's health deterioration can take many years to manifest ally affect its mortality (Kotzamanis, Zafeiris, and Kostaki 2022). Economic crisis was associated with remarkable adverse effects on perinatal outcomes and infant mortality, mainly explained by long-term unemployment and income reduction (Siahanidou et al. 2019). It is also important to note that the economic crisis itself was precipitated in part with inaccurate data reporting (Rauch et al. 2011), the notable absence of an assessment of the quality of routine service statistics reporting, and the unlikeliness that this reporting was immune from the severe budget cuts.

### 2.5 The male disadvantage

According to the United Nations Inter-agency Group for Child Mortality Estimation, child mortality is more common for boys in all countries (You et al. 2015). Several studies have further documented higher male perinatal and infant mortality (Ulizzi and Zonta 2002; Lawn, Cousens, and Zupan 2005; Sawyer 2012; Naeye et al. 1971; Carlsen, Grytten, and Eskild 2013; Fottrell et al. 2015). Girls have a well described biological survival advantage in the neonatal period. They are less vulnerable to birth complications and infections and have fewer inherited abnormalities. All else being equal, Thus, the ratio of infant mortality among boys to infant mortality among girls is greater than one, provided both sexes have equal access to food and medical care (Sawyer 2012).

## 3 Identification strategy

Previous studies use a variety of methods including life expectancy analysis, interrupted time series, time trends, and pre-post comparisons. We applied Synthetic Control Method (SCM) which relaxes the parallel trend assumption to build a more flexible latent factor model where the counterfactual is projected from the extant mortality attributes of Greece before austerity measures. Moreover, SCM enables a direct estimate of excess mortality attributable to the study event.

### 3.1 Setup

This study evaluates the effect of austerity policies by estimating the missing counterfactual realization of the mortality dynamics and simulating its trajectory in the hypothetical absence of the austerity policy and designated cuts in public healthcare. We apply the synthetic control estimator (Abadie, Diamond, and Hainmueller 2010; Kreif and DiazOrdaz 2019; Bonander 2021) to estimate the effect of austerity on infant mortality rate in Greece. Through the application of the potential outcomes framework (Rubin 1973), the infant mortality trajectories of Greece can be simulated through the combination of attributes of other countries that have similar characteristics but have not undergone a severe austerity during the study period. This approach allows the construction of a counterfactual representation of Greece by making use of weighted averages reflecting the resemblances of infant mortality characteristics before the introduction of austerity policies.

It should be noted that the impact of austerity policies is reflected by the infant mortality difference between Greece and the estimated counterfactual absent deep public health spending cuts in the post-intervention period. The synthetic control group that best reproduces infant mortality attributes of Greece prior to the austerity policies allows us to predict and infer the plausible level of mortality rate if the austerity cuts in public health spending were hypothetically never implemented. Provided that the synthetic control group is not tainted by the presence of similarly harsh spending reductions, the outcome difference plausibly reflects the effect of austerity policies on infant mortality rate.

Suppose that we observe $J + 1$ countries over $t = 1,2,...T$ period where $J$ denotes the country affected by a deep reduction in public health spending and $\{2,...J + 1\}$ represent the donor pool of OECD countries not affected by the intervention. As a mimic of the treatment, austerity policy occurs at time $T_0$ and lasts in the post-intervention period so that $t < T_0 < T$. Our aim is to estimate the impact of the health-related austerity on infant mortality as the underlying outcome variable. Without the loss of generality, let $Y_{i,t}^N$ denote the outcome for country $i$ at time $t$ in the absence of the austerity health policy. By contrast, let $Y_{i,t}^I$ be the outcome for country $i$ that would be observed if it were exposed to the austerity. By assuming that the austerity has no effect on the outcome of interest in the pre-intervention period, it follows that $Y_{i,t}^N = Y_{i,t}^I$ for all $i$ and $t < T_0 + 1$ and we further assume that a harsh form of austerity in health care only affects Greece. For each $t > T_0$, the effect of the austerity on infant mortality can be written as follows:

$$\theta_{1,t} = Y_{1,t}^I - Y_{1,t}^N = \underbrace{Y_{1,t}}_{\text{observed}} - \underbrace{Y_{1,t}^N}_{\text{counterfactual}} \tag{1}$$

where the key challenge is to construct the unobserved counterfactual and estimate $\theta$ accordingly. To reconstruct the missing counterfactual scenario, we rely on the latent factor model that accommodates pre-$T_0$ infant mortality path to estimate the unobserved counterfactual component:

$$Y_{i,t}^N = \eta_t + \pi_t \cdot Z_i + \mu_t \cdot \phi_i + \varepsilon_{i,t} \tag{2}$$

where $\eta_t$ is an unobserved common factor that mimics time-fixed effects, $Z_i \in \mathbb{R}^r$ is a vector of observed time-varying and time-invariant covariates unaffected by the civil war, $\pi_t' \in \mathbb{R}^r$ is a vector of known parametric factor loadings, and $\mu_t' \in \mathbb{R}^F$ is a vector of common unobserved factors, and $\phi_t' \in \mathbb{R}^F$ is a vector of unknown factor loadings. The term $\varepsilon_{i,t}$ denotes transitory outcome shocks with $\varepsilon_{i,t} \sim \mathbb{N}(0,1)$ form of distribution. The key advantage of the latent factor model is to allow heterogeneous response of outcomes to multiple unobserved factors ($\mu_t \cdot \phi_i$) and embeds time trend models therein. The proposed latent factor model implicitly assumes that the number of common unobserved factors ($\mu_t$) is fixed over time which invokes the absence of structural breaks. The basic intuition behind the latent factor model is to reweigh the control group so that the synthetic version of Greece will match its observed $Z_i$. Hence, $\phi_i$ will be matched by default and the set of unobserved common factors will not be projected out of the outcome generating process allowing for heterogeneous response to multiple unobserved factors. Consider a simple $J \times 1$ single dimensional vector of weights $W = (w_2, \dots, w_{J+1})'$ where $w_j \geq 0$ for $j = 2, \dots, J+1$ and $\sum_{j=2}^{J+1} w_j = 1$. Notice that each particular value of $W$ represents the weighted average of the implicit mortality attributes of the countries from a donor pool where $w_j \neq 0$. Let **X** represent pre-$T_0$ mortality trajectory combined into a single vector that is partitioned into $\mathbf{X}_{Greece}$ and $\mathbf{X}_0$ where the former captures the values of matching variables for Greece and $\mathbf{X}_0$ represents the values of the variables for the countries unaffected by the austerity. To appropriately measure the discrepancy between Greece and the unaffected countries, we estimate $W$ through a nested optimization method using constrained quadratic programming routine to find best-witting set of weights conditional on the matrix of covariates. It is based on the algorithm using the Vanderbei (1999) interior point method to solve the constrained quadratic programming problem, and is implemented via C++ plugin with 5 percent margin for the constraint violation tolerance using high-speed maximum likelihood approach as our default. We use a standard inner optimization to minimize a simple Euclidean distance between $\mathbf{X}_{Greece}$ and $\mathbf{X}_0$ to find the best-fitting weight set:

$$W^* = \underset{W}{\operatorname{argmin}} \|\mathbf{X}_{Greece} - \mathbf{X}_0 W\|_V = \sqrt{(\mathbf{X}_{Greece} - \mathbf{X}_0 W)' \mathbf{V}(\mathbf{X}_{Greece} - \mathbf{X}_0 W)} \quad (3)$$

where **V** is a symmetric diagonal matrix with positive components wherein $k$ diagonal elements $(v_1, \ldots v_k)$ represent the predictive weights of the pre-$T_0$ fitted variables. In the next step, in the outer optimization, **V** can be estimated to minimize mean square prediction error in the pre-austerity period such that:

$$V^* = \underset{V}{\operatorname{argmin}} \big(\mathbf{Y}_{Greece} - \mathbf{Y}_0 \mathbf{W}^*(\mathbf{V})\big)' \big(\mathbf{Y}_{Greece} - \mathbf{Y}_0 \mathbf{W}^*(\mathbf{V})\big) \quad (4)$$

where $\mathbf{Y}_{Greece}$ denotes pre-austerity infant mortality rate of Greece and $\mathbf{Y}_0$ captures a variety of linear combinations of pre-austerity mortality rate of the unaffected countries that represent potential synthetic control units. To ensure that optimal non-negative weights exist as a convex linear combination of control countries, it is necessary that the interpolation biases are not present (Abadie 2021). Under the regularity of these conditions, the average treatment effect of austerity policy on infant mortality can be written as:

$$\hat{\theta}_1 = \frac{1}{(T-T_0)} \cdot \sum_{t>T_0} \left(Y_{1t} - \sum_{j=2}^{J+1} w_j^* Y_{j,t}^N\right) \tag{5}$$

Furthermore, the synthetic representation of Greece in the pre-austerity period reinforces the estimated unobserved counterfactual from the latent factor provided that the set of weights is identified in the inner and outer optimization which implies that the outcome based on reweighting the synthetic control group is as follows:

$$Y_{W,t} = \sum_{j=2}^{J+1} w_j \cdot Y_{j,t} = \eta_t + \pi_t \cdot \left(\sum_{j=2}^{J+1} w_j Z_j\right) + \mu_t \cdot \left(\sum_{j=2}^{J+1} w_j \phi_j\right) + \left(\sum_{j=2}^{J+1} w_j \varepsilon_{j,t}\right) \tag{6}$$

provided that pre-austerity period is sufficiently large and that interpolation bias issues are properly mitigated, the synthetic control estimator provides a plausible representation of the counterfactual scenario in response to the austerity under time-varying heterogeneity. It should be noted that we divide the pre-intervention period into the training and validation period to choose **V** such that the resulting synthetic control unit can plausibly approximate the outcome trajectory before the austerity policy. Provided that the number of pre-austerity periods is large enough, for a given **V**, **W** can be computed directly using the covariate matrix from the training period to minimize the mean squared prediction error during the validation period when the set of weights is produced.

### 4.2  Inference

Since large-sample asymptotic inference is not possible with synthetic control method, Abadie et. al. (2010) advocate the use of treatment permutation method for inference on the treatment effect of interest (Abadie, Diamond, and Hainmueller 2010). In our setup, we undertake a simple treatment permutation in space to detect whether the effect of the austerity measures on infant mortality is statistically significant at conventional levels. Treatment permutation consists of the iterative assignment of the austerity shock to the unaffected countries and use the outcome gaps to build the appropriate distribution of placebo effect. We proceed in two steps. In the first step, we estimate the placebo treatment effect by assigning the 2009 austerity package to each unaffected country in the OECD donor pool. In the second step, we compute the fraction of countries having the placebo effect at least as large as that of Greece. Our intuition is simple and straightforward. If the effect of austerity is both negative and statistically significant, the outcome gap should be unique to Greece and not perceivable elsewhere in the donor pool. By contrast, if the effect of austerity is not statistically significant, the placebo distribution should indicate no difference in the estimated effect of austerity for Greece and the placebo runs. Therefore, in step two, we calculate the empirical p-value for effect of austerity package on infant mortality in Greece from a simple two-tailed empirical distribution:

$$\mathbb{p}_{i,t} = \frac{\sum_{j=2}^{J+1} \mathbf{1} \cdot \{\hat{\theta}_{j \in J+1} \geq \hat{\theta}_1\}}{J}$$

where $\mathbb{p}_{i,t}$ denotes two-tailed empirical p-value for the estimated dynamic treatment effect of austerity, $\hat{\theta}_{j \in J+1}$ denotes the full-sample placebo effect, $\hat{\theta}_1$ designates the estimated treatment effect of war on Greece, and $J$ indicates the size of the donor pool. Notice that the empirical p-value represents the probability to obtain dynamic average treatment effect at least as large as the one for Greece. However, if the prediction accuracy in the placebo simulation is low, the null hypothesis on the treatment effect of civil war is prone to over-rejection given a relatively large rarity of obtaining a large placebo effect. To address the type-II error, we compute pre-austerity average prediction error and constrain the inference procedure to ensure that p-values are estimated only if the prediction error in each placebo simulation is lower or equal to the mean prediction error for Greece. In addition, we discard the countries with mean prediction error at least four times that of Greece to partially eliminate the possibility of under-rejecting the null hypothesis that may arise if poorly fit placebos were included in the simulation. Our intuition behind the intertemporal behaviour of empirical p-values is two-fold. First, if the accuracy-adjusted p-values are consistently low from the early years of austerity onward, the analysis may highlight the austerity as a source of persistent and elevated increase in infant mortality. And second, if the adjusted p-values by the end of the sample are high, then the notion of austerity as a temporary negative shock with few long-term implications becomes more salient. Albeit imperfect by default, such inference procedure allows us to detect whether the austerity policy has led to a permanent deterioration on infant mortality trajectories to evaluate the significance of the effect in greater detail.

## 4 Data and samples

### 4.1 Outcomes

The primary outcome variable of interest is infant mortality rate. Using the estimates from United Nations Demographic Yearbook, we approximate the mortality rate as the total number of deaths in a given year of female or male children less than one year of age, divided by the total number of female or male live births in the same year, multiplied by 1,000. The rate of mortality is the approximation of the number of deaths per 1,000 children born alive who die within one year of birth. The series is calculated from the estimates and projections of the number of child survivors at the age of 1 disaggregated by gender. Since mortality rates for different age groups comprise important indicators of health status and social development, rate of infant-related death rate invariably indicates the mortality level among the most vulnerable age group within the population. Specifically, we consider gender-disaggregated mortality rate and compare the infant mortality trajectories between Greece and its donor pools both for boys and girls to better understand gender-specific effect disparity behind fiscal austerity measures.

Figure 1 depicts overall infant mortality trajectories in Greece and OECD countries for the period 1991-2020. A broad comparison of trajectories shows that prior to the onset of the fiscal austerity measures, Greek infant mortality trajectory appears to be substantially below the OECD average. Two detailed patterns emerge after the implementation of austerity measures. First, the austerity measures appear to have push the mortality trajectory of Greece closer to the OECD average and have decimated noteworthy and substantial public health gains and advantage ahead of OECD peers from 1990 onwards. And second, the infant mortality rate does not seem to have slowed down but has increased in absolute terms after the austerity measures were introduced. By the beginning of 2016, Greek infant mortality trajectory evolves in tandem with the OECD average prior to the onset of the COVID-19 pandemic.

**Figure 1**: Infant mortality trajectories in OECD and Greece, 1991-2020

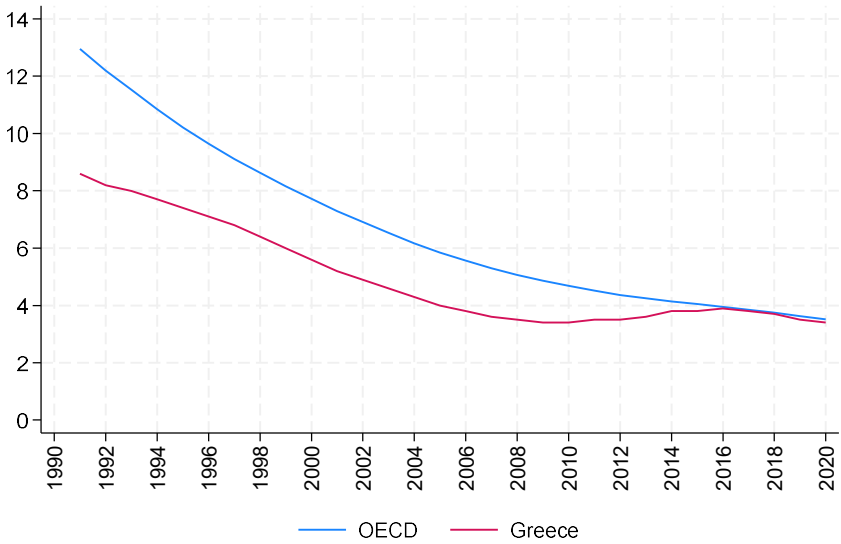

Table 1 reports pre-austerity infant mortality imbalance between Greece, its synthetic control group and the overall donor pool of OECD member states based on the full-outcome path nested dynamic optimization of the Greece and its convex hull of mortality attributes of other OECD members. The comparison of outcome balance show that in the absence of convex optimization and synthetic matching between Greece and the OECD donor pool, a simple comparison of pre-austerity mortality trajectories may not be able to uncover the overall mortality effect associated with the austerity measures. The overall bias between Greece and the OECD donor pool in the absence of the convexity-based analysis is between 26 percent and 56 percent higher than the observed mortality rate values for Greece. By contrast, the estimated bias between Greece and its synthetic control group in the pre-austerity period is between 0.13 percent and 1.75 percent, respectively. Using full outcome path optimization to seek exact mortality attributes of OECD member states that fall within the convex hull of Greek mortality dynamics nearly erases estimated bias from the comparison. It should be noted that low bias does not appear to be driven by disproportionately highly-leverage pre-austerity mortality rate in the benchmark years since the relative weight of each benchmark mortality rate is relatively evenly distributed across the pre-austerity period, and, as such, does not pose an unresolved caveat as a stumbling block against valid inference on the treatment effect of the austerity measures. Figure 2 graphically depicts standardized bias in the mortality rate comparison across the full span of benchmark years in the pre-austerity period between Greece, its synthetic control group and the OECD member state donor pool. The comparison indicates that the synthetic control group from the OECD donor pool provides an excellent quality of the fit with the Greek infant mortality trajectory in the pre-treatment period.

Table 1: Pre-austerity outcome path imbalance

|  | $\hat{\mathbf{V}}$ | $\mathbf{X}_{t<T_0}^{Greece}$ | Synthetic Control Group | | Overall Donor Pool | |
|---|---|---|---|---|---|---|
|  |  |  | $\mathbf{X}_{t<T_0}^{0}$ | Bias | Mean | Bias |
| Infant Mortality$_{t=1991}$ | 0.061 | 8.60 | 8.52 | -0.89% | 11.62 | 35.16% |
| Infant Mortality$_{t=1992}$ | 0.059 | 8.20 | 8.31 | 1.40% | 11.11 | 35.49% |
| Infant Mortality$_{t=1993}$ | 0.065 | 8.00 | 8.01 | 0.13% | 10.58 | 32.30% |
| Infant Mortality$_{t=1994}$ | 0.114 | 7.70 | 7.71 | 0.18% | 10.04 | 30.38% |
| Infant Mortality$_{t=1995}$ | 0.007 | 7.40 | 7.38 | -0.22% | 9.52 | 28.66% |
| Infant Mortality$_{t=1996}$ | 0.004 | 7.10 | 7.07 | -0.38% | 9.05 | 27.46% |
| Infant Mortality$_{t=1997}$ | 0.055 | 6.80 | 6.74 | -0.94% | 8.61 | 26.63% |
| Infant Mortality$_{t=1998}$ | 0.104 | 6.40 | 6.37 | -0.49% | 8.22 | 28.45% |
| Infant Mortality$_{t=1999}$ | 0.143 | 6.00 | 5.98 | -0.27% | 7.84 | 30.75% |
| Infant Mortality$_{t=2000}$ | 0.051 | 5.60 | 5.61 | 0.21% | 7.51 | 34.07% |
| Infant Mortality$_{t=2001}$ | 0.065 | 5.20 | 5.24 | 0.80% | 7.17 | 37.96% |
| Infant Mortality$_{t=2002}$ | 0.027 | 4.90 | 4.89 | -0.24% | 6.88 | 40.33% |
| Infant Mortality$_{t=2003}$ | 0.117 | 4.60 | 4.59 | -0.11% | 6.61 | 43.59% |
| Infant Mortality$_{t=2004}$ | 0.045 | 4.30 | 4.33 | 0.59% | 6.33 | 47.31% |
| Infant Mortality$_{t=2005}$ | 0.017 | 4.00 | 4.02 | 0.48% | 6.09 | 52.17% |
| Infant Mortality$_{t=2006}$ | 0.001 | 3.80 | 3.84 | 1.04% | 5.85 | 53.95% |
| Infant Mortality$_{t=2007}$ | 0.039 | 3.60 | 3.66 | 1.59% | 5.63 | 56.29% |
| Infant Mortality$_{t=2008}$ | 0.003 | 3.50 | 3.48 | -0.62% | 5.42 | 54.89% |
| Infant Mortality$_{t=2009}$ | 0.025 | 3.40 | 3.34 | -1.75% | 5.23 | 53.79% |

Notes: the table reports pre-austerity infant mortality trajectory imbalance between Greece, its synthetic control group and the overall donor pool of OECD member states. The matrix $\hat{\mathbf{V}}$ denotes the set of normalized variable weights in the diagonal and $\mathbf{X}_{t<T_0}^{Greece}$ represents the mortality rates of Greece in each pre-austerity period. The table also reports the corresponding bias between Greece and its synthetic control group, and Greece versus the overall donor pool of OECD member states

**Figure 2**: Standardized pre-austerity outcome path imbalance

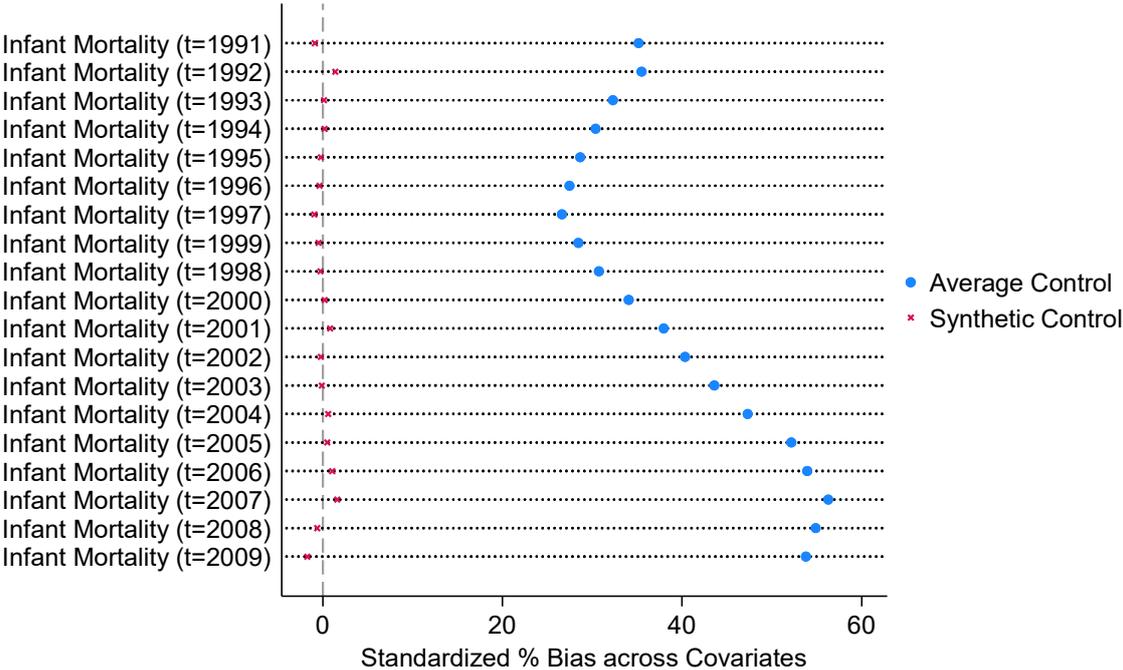

Figure 3 details the composition of the synthetic control group that best reproduces the pre-austerity infant mortality trajectory of Greece but has not been affected by the deep public health spending cuts. Therefore, the overall infant mortality of Greece in the pre-austerity period is best synthesized by the implicit convex attributes of Iceland (49 percent), Estonia (19 percent), Portugal (13 percent), Ireland (7 percent), Latvia (5 percent), and Costa Rica (4 percent), respectively. The infant mortality trajectories of other OECD member states fall outside the convex hull of Greek mortality characteristics and therefore do not receive a positive weight in the validation stage. Since none of the OECD donor countries with positive weight had undergone such drastic reduction of public health expenditure in response to a foreign intervention by Troika, it is unlikely that the donor pool is tainted by the confounding shocks that could be perceptible elsewhere and, thus, violate SUTVA (i.e. stable unit treatment value assignment) assumption. Although each individual donor pool was affected by the economic downturn in the same period as Greece, public health fiscal reforms and spending cuts in Greece has been rather unique both in terms of magnitude, severity and duration.

**Figure 3**: Composition of the synthetic control group for overall infant mortality

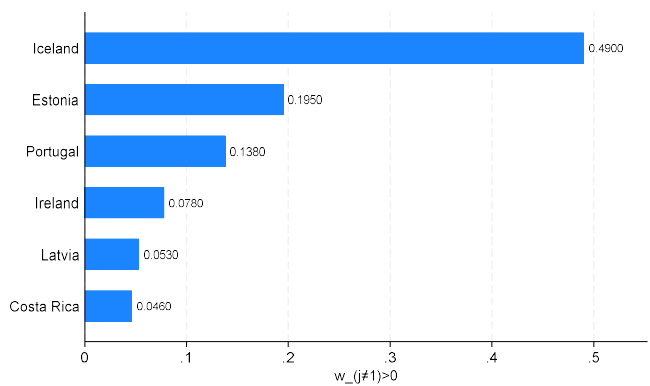

**5    Results**

*5.1    Baseline*

This study finds substantial negative impact of the austerity policies, characterized by immediate and persistent increase in infant mortality rate relative to the counterfactual scenario. Figure 4 reports the estimated infant mortality effects of the fiscal austerity measures in Greece. More specifically, the figure reports infant mortality trajectories of Greece and its synthetic control group in the pre-austerity period (1991-2009) and post-austerity period (2010-2020). The right part of the figure plots year-specific pre- and post-austerity effect indicating the change in infant mortality rate. The evidence shows that the synthetic control estimator provides an excellent quality of the fit between Greece and its artificial control group prior to the imposition of the austerity measures with almost zero discrepancy. Based on the existing benchmark of the quality of fit criteria (Adhikari and Alm 2016), the estimated root mean squared prediction error (RMSE = 0.044) is less than 1 percent of the pre-treatment error margin, and indicates a very low degree of infant mortality path imbalance between Greece and its synthetic control group in the pre-austerity period.

The estimated synthetic counterfactual scenario invariably suggests that the austerity measures promulgated an immediate and substantial increase in the infant mortality rate that appears to be relatively large and persistent up to the end of post-austerity period. In quantitative terms, the average treatment effect of austerity-induced fiscal spending reduction in public health care is associated with 850 additional infant deaths for each year of the austerity policies relative to the synthetic OECD counterfactual. Our synthetic control estimates indicate that the estimated mortality toll is around 210 additional infant deaths in the first year of austerity, and gradually increases up to 1,210 additional deaths by 2017 whilst decreasing to 970 additional deaths by the last post-austerity year before the COVID-19 pandemic. The cumulative infant mortality toll behind the austerity policies is estimated at 9,360 infant deaths in the period 2010-2020 that can be potentially attributed to deep fiscal expenditure reduction in public health care. Table 2 reports the full post-treatment distribution of overall and gender-disaggregated infant mortality effect associated with the austerity policies.

**Figure 4**: Infant mortality effect of austerity measures in Greece, 1991-2020

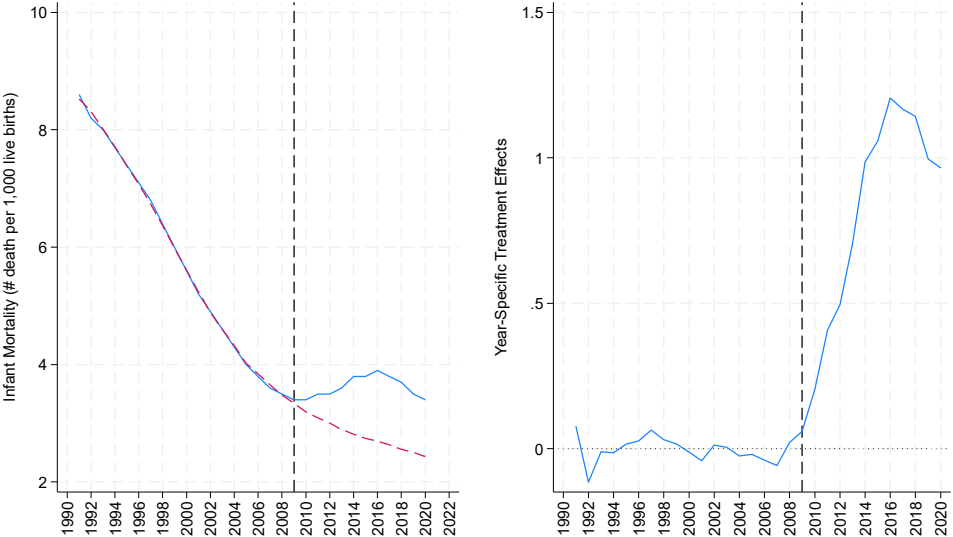

The estimated infant mortality increase in response to the austerity policies is characterized by a noteworthy gender-specific variation. Figure 5 reports gender-disaggregated infant mortality effect of austerity measures in Greece. The evidence suggests that the mortality increase was disproportionately more strongly concentrated among boys. By way of example, the estimated average treatment effect of austerity policies on infant mortality is around 29 percent higher for boys. The disparity in the estimated effects appears to be relatively uneven throughout the post-austerity period. In the first year of austerity policies, the mortality rate for boys is around 10 percent higher compared to the girls' mortality rate. By the peak of the mortality rate increase in 2016, the disparity tends to increase to 30 percent more disproportionate impact for boys. By the end-of-sample period prior to the onset of COVID19 pandemic, the post-treatment effect disparity is around 32 percent. Without the loss of generality, gender-specific synthetic control estimates invariably suggests that the overall infant mortality effect of the fiscal austerity measures is around one third higher and stronger for boys which have been substantially more disproportionately affected by the adverse capacity constraints on public health and birth care through increased unmet medical needs, staff shortage and equipment shortfall.

**Figure 5**: Gender-disaggregated infant mortality effect of austerity measures in Greece, 1990-2020

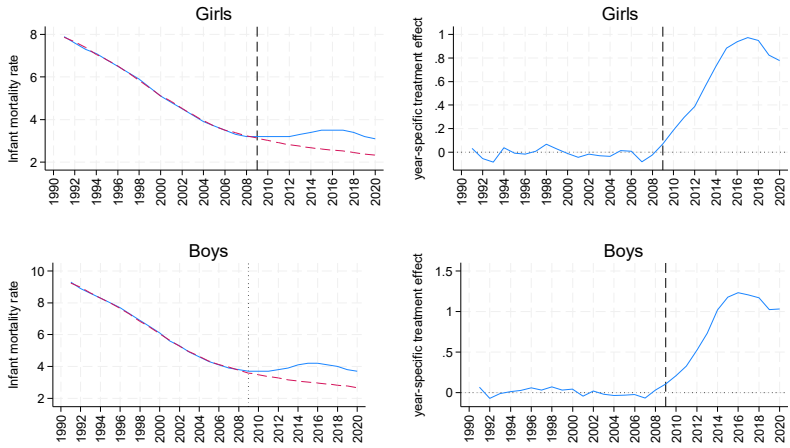

**Figure 6**: Gender-disaggregated synthetic control groups

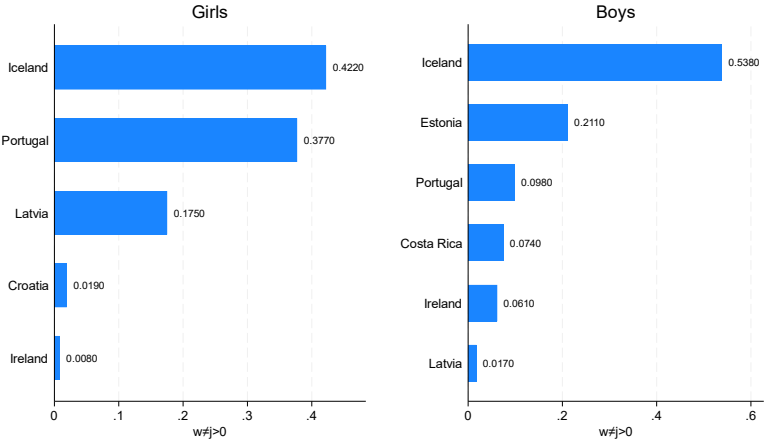

Figure 6 illustrates gender-disaggregated synthetic control groups that best reproduce the infant mortality trajectories of Greece in the pre-austerity period but have, at the same, not been hit by such deep and adverse fiscal shock and spending reduction in public health care. For instance, girls infant mortality rate of Greece prior to the imposition of the austerity measures is best reproduced by the convex combination of mortality dynamics and attributes of Iceland (42 percent), Portugal (37 percent), Latvia (17 percent), Croatia (2 percent) and Ireland (<1 percent), respectively. Furthermore, the boys infant mortality trajectory prior to the austerity period is best synthesized by the convex combination of mortality attributes of Iceland (54 percent), Estonia (21 percent), Portugal (9 percent), Costa Rica (7 percent), Ireland (6 percent), and Latvia (2 percent). The comparison confirms a very strong similarity with the baseline synthetic control group reported in Figure 3 and does not indicate a source of false comparison or a distinctive composition of the donor pool that could potentially render the estimated infant mortality gap incomprehensible.

**Table 2**: Post-austerity outcome balance and treatment effect composition

|      | Overall | | | Girls | | | Boys | | |
|------|---------|---|---|-------|---|---|------|---|---|
|      | Actual Infant Mortality Rate | Synthetic Infant Mortality Rate | Treatment Effect | Actual Infant Mortality Rate | Synthetic Infant Mortality Rate | Treatment Effect | Actual Infant Mortality Rate | Synthetic Infant Mortality Rate | Treatment Effect |
| 2010 | 3.40 | 3.19 | 0.21 | 3.20 | 3.01 | 0.19 | 3.70 | 3.49 | 0.21 |
| 2011 | 3.50 | 3.09 | 0.41 | 3.20 | 2.91 | 0.29 | 3.70 | 3.37 | 0.33 |
| 2012 | 3.50 | 3.01 | 0.49 | 3.20 | 2.81 | 0.39 | 3.80 | 3.28 | 0.52 |
| 2013 | 3.60 | 2.89 | 0.71 | 3.30 | 2.74 | 0.56 | 3.90 | 3.17 | 0.73 |
| 2014 | 3.80 | 2.81 | 0.99 | 3.40 | 2.67 | 0.73 | 4.10 | 3.08 | 1.02 |
| 2015 | 3.80 | 2.74 | 1.06 | 3.50 | 2.62 | 0.88 | 4.20 | 3.02 | 1.18 |
| 2016 | 3.90 | 2.69 | 1.21 | 3.50 | 2.56 | 0.94 | 4.20 | 2.97 | 1.23 |
| 2017 | 3.80 | 2.63 | 1.17 | 3.50 | 2.52 | 0.98 | 4.10 | 2.90 | 1.20 |
| 2018 | 3.70 | 2.56 | 1.14 | 3.40 | 2.45 | 0.95 | 4.00 | 2.83 | 1.17 |
| 2019 | 3.50 | 2.50 | 1.00 | 3.20 | 2.38 | 0.82 | 3.80 | 2.78 | 1.02 |
| 2020 | 3.40 | 2.43 | 0.97 | 3.10 | 2.32 | 0.78 | 3.70 | 2.67 | 1.03 |
| Mean | 3.63 | 2.79 | 0.85 | 3.32 | 2.64 | 0.68 | 3.93 | 3.05 | 0.88 |

*5.2    Robustness checks*

*5.2.1    In-space placebo analysis*

To estimate the statisistical significance of the gap between Greece and its synthetic peer and validate inference on the effect of austerity measures we iteratively assign the austerity policies to the donor pool of countries. Then, the synthetic control estimator is applied to each OECD member state from the donor pool to build the in-space placebo distribution of the infant mortality gap. The intuition behind such comparison is both simple and straightforward. If the placebo distribution indicates that the infant mortality gap in the OECD member states is comparable with the actual mortality gap in Greece, then our interpretation is that it is unlikely that our estimates provide evidence of the statistically significant infant mortality effect associate with austerity measures. By contrast, if the estimated infant mortality gap is unique to Greece and substantially distinctive from the placebo distribution, the notion of statistically significant infant mortality effect becomes more credible and plausible. Our in-space placebo analysis proceeds in two steps. First, the ratio of post- and pre-austerity RMSE is compared. If the infant mortality gap is unique to Greece and can be only scarcely perceived elsewhere, the error ratio for Greece should be high and no other country from the donor pool should be reach such high threshold in the distribution. And second, based on the comparison of post- and pre-austerity RMSE parameter, we compute the quasi p-value on the null hypothesis behind infant mortality gap and determine the probability that the estimated infant mortality effect happened by chance for each year in the post-austerity period. If the proportion of countries characterized by the similar post-/pre-austerity RMSE ratio as Greece is high, then the notion of statistical significance of the effect would be limited. Conversely, if the proportion of countries from the donor pool having similar RMSE ratio to Greece is low, then the null hypothesis can be more easily rejected and the notion that the effect is unique to Greece and cannot be perceived elsewhere in the donor pool becomes more readily acceptable. Figure 7 compares post- and pre-austerity RMSE ratio of Greece against the OECD donor pool. It becomes apparent that Greece is characterized by the highest post- vs. pre-austerity RMSE ratio in the full sample of OECD countries. Not a single OECD country tends to approach the RMSE ratio of Greece which implies that the notion of effect uniqueness can only seldom be disputed. It should be noted that Greece is also characterized by the highest post- vs. pre-austerity ratio either when the infant mortality effect is estimated separately for boys and girls.

**Figure 7**: Comparison of post and pre-austerity RMSE ratio

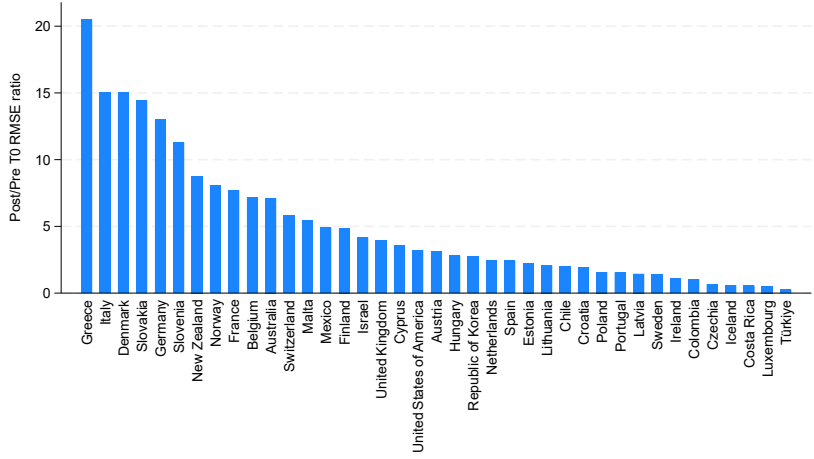

Figure 8 presents the intertemporal distribution of quasi p-values on the null hypothesis behind the infant mortality effect of austerity measures. The horizontal axis exhibits the length of the post-austerity period and designated each consecutive post-austerity year whilst the vertical axis reports the probability that the estimated post-austerity infant mortality gap happened by chance. In evaluating the statistical significance of the infant mortality gaps, our decision rule is to reject the null hypothesis is the quasi p-value for each post-austerity period is within the conventional 10 percent threshold. The figure presents the intertemporal distribution of quasi p-values for the overall mortality specification as well as for gender-disaggregated infant mortality effects. A possible caveat against a valid inference on the post-treatment effect of austerity policies on infant mortality arises from a large density of placebo runs with a poor quality of the fit which seldom provide the necessary information to measure the relative rarity of estimating a large post-austerity infant mortality gap for a country that was otherwise well fitted prior to the austerity policies. We partially address this issue and discard those countries from the placebo simulation that have the estimated pre-austerity MSPE more than twice the MSPE of Greece. By avoiding overly lenient cutoff, our inference is based on a relatively parsimonious and conservative specification focussing on the countries that can fit almost as well as Greece in the pre-austerity period. The evidence suggests that the null hypothesis behind the treatment effect of austerity measures on infant mortality can be relatively easily rejected across the entire post-austerity period. The quasi-p-values from the overall infant mortality specification are consistently around 0.000 for the full post-austerity period. The p-value on the null hypothesis drops to zero in the second year of the austerity policies, indicating that statistically significant increase in infant mortality unfolded in 2011. The p-value does not tend to disappear outside 5 percent or 10 percent significance threshold for the entire duration of the post-treatment period, which confirms statistically significant and permanent effect of austerity measures on infant mortality, and suggests that austerity policies led to a long-lasting upward derailment of the Greek infant mortality trajectory.

**Figure 8**: Inference on the infant mortality effect of austerity measures

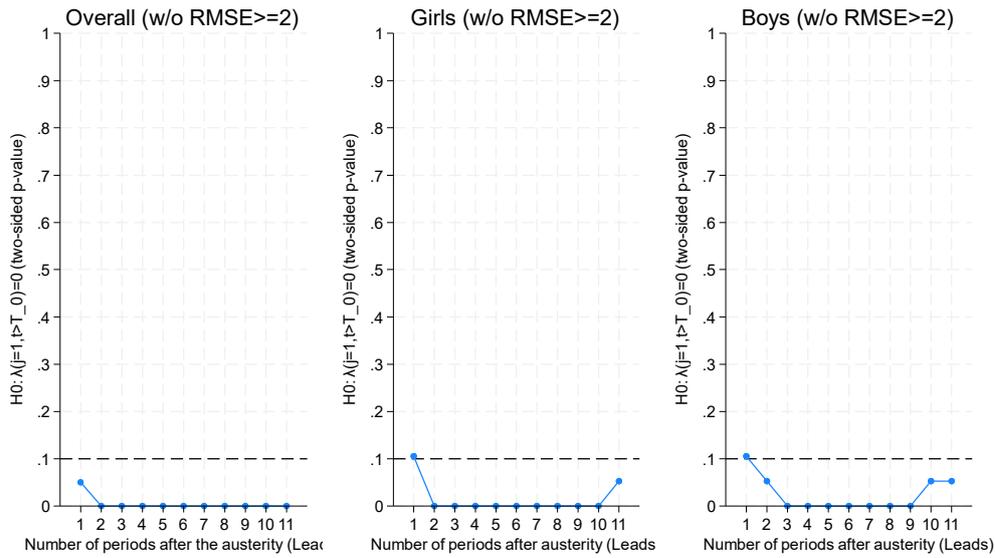

### 5.2.2 In-time placebo analysis

Another potential objection against the internal validity of the treatment effect of austerity measures on infant mortality arises from the possible anticipation of the effect or confounding influence of the pre-austerity policies and changes that could contaminate the treatment effect with the component of the shock that is distinctive from the imposition of austerity measures. Such shocks could include pre-austerity health care reforms, preventive care or fiscal adjustment and stimulus measures geared towards the public health system and service capacity. Under such conditions, the synthetic control would estimate large mortality effects when applied to the deliberately wrong dates of austerity policy measures. We address this particular caveat and perform in-time placebo analysis by backdating the austerity policy to the middle of the pre-treatment period where no austerity measures were implemented.

Figure 9 reports in-time placebo analysis of the overall and gender-disaggregated infant mortality effect of the austerity measures deliberately using the wrong year of the policy intervention. The evidence suggests that the synthetic control estimator provides an excellent quality of the fit between Greece and its synthetic peer in the years prior to the quasi date of the policy intervention both for the overall and gender-specific mortality rates. By contrast, our in-time falsification analysis shows that real and synthetic trajectories of mortality do not diverge after the placebo year. Instead, the mortality trajectories tend to diverge immediately and rapidly after the true date of the austerity-related intervention in 2009 across the entire spectrum of mortality specifications. Thus, arbitrarily assigning the austerity policy to the wrong year, provides further evidence that the infant mortality increase is a result of the austerity measures.

**Figure 9**: In-time placebo analysis of the infant mortality effect of austerity measures in Greece, 1991-2020

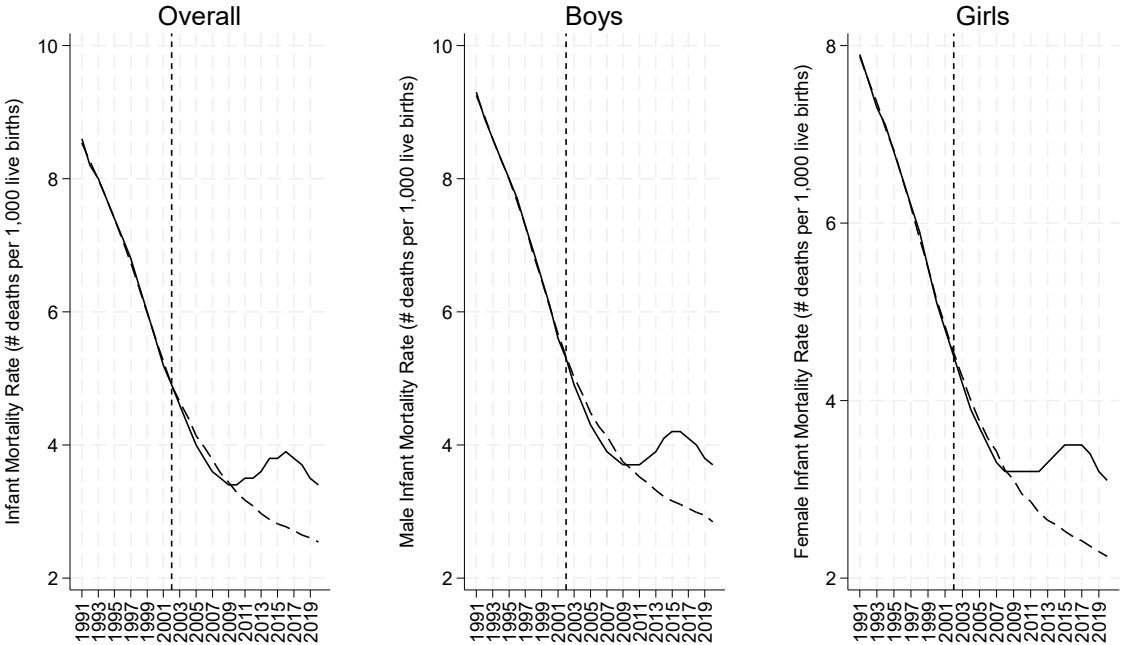

### 5.2.3 Differential trend analysis

One of concerns arising from the estimated infant mortality effect of austerity measures hinges on the nature of the shock. Given that the infant mortality gap between Greece and each variant of its synthetic control group unfolds immediately and does not disappear up to the present day, the question that arises is whether the austerity measures induced a permanent and differential change in infant mortality rate. In the absence of a permanent differential change, the notion of long-lasting impacts of the austerity measures would be questionable as the absence of the break would have indicated a more transitory and temporary effect of austerity measures on the mortality rate. A decline of infant mortality both in Greece and its OECD-level synthetic control group after 2015 may posit abundant intuitive evidence in support of the temporary effect of austerity policies on infant mortality. Conversely, if our evidence corroborates the notion of a reasonably strong upward structural break in infant mortality trajectory that can be properly differentiated from the post-intervention trend in the synthetic control groups, the notion of the more permanent effect of austerity policy under such conditions becomes more credible. To assess whether the austerity policies proliferated a permanent or temporary increase in infant mortality we apply the differential trend assumption test. This approach as been applied to test the differential effect of the trans-fats ban on cardiovascular mortality rate in Denmark (Spruk and Kovac 2020), and further extended into a variety of applications ranging from the human development effects of civil wars (Kešeljević and Spruk 2024) and long-term effects of oil discoveries (Gilchrist et al. 2023).

Testing the differential trend assumption in the single-treatment setup of synthetic control estimator is both simple and intuitive. If the austerity measures triggered a temporary and weak infant mortality effect, the mortality gap between Greece and synthetic control group in the post-intervention period relative to the period before the intervention should be negligible and statistically insignificant at conventional levels. Conversely, if the austerity polic produced a permanent and reasonably strong break in the trend of mortality, the respective mortality gap between Greece and its synthetic counterparts should differ both markedly and statistically significantly in the post-austerity period whilst zero gap should be detectable in the pre-austerity period if the quality of the fit in the designated period is good. The null hypothesis behind the differential trend assumption can be tested by computing the supremum test statistics on the triple-difference interaction coefficient in the semi-difference-in-differences regression of mortality rate gap on the indicator of Greece and post-austerity period in the presence of both observed and unobservable confounders. To address the potential endogeneity between the austerity measures and infant mortality rate, we include two lags of the mortality gap in each semi-DiD regression specification to partially rule out the reverse causation channel from the inference.

Table 3 reports the differential trend assumption test on the change in the gap of infant mortality between Greece and its synthetic control group in the post-austerity period relative to the pre-austerity benchmark period. Under the null hypothesis, the mortality gap between Greece and its synthetic control group exhibits no structural break and is not characterized by statistically significant and perceptible structural break in the post-austerity period. The evidence suggests that the null hypothesis of no differential structural break in mortality gap can be rejected at 1% significance threshold. Using either full mortality specification or the gender-disaggregated ones, we find evidence of strong and statistically significant upward break in the mortality trajectory of Greece relative to its synthetic control group (i.e. p-value = 0.000). This supports the finding that the austerity policies did not lead to a mere temporary increase in infant mortality but instead promulgated a long-term effect. . Figure 10 depicts post- vs. pre-austerity trends in the mortality gap between Greece and its synthetic control group alongside the 95% confidence intervals.

Table 3: Testing differential trend assumption behind infant mortality trajectories

|  | Full distribution | Girls | Boys |
|---|---|---|---|
| $\chi^2$ test statistics | 31.17 | 45.08 | 43.81 |
| (p-value) | (0.000) | (0.000) | (0.000) |
| Exogenous variable #1 | $\Delta Y_{j=Greece, t-1}$ | $\Delta Y_{j=Greece, t-1}$ | $\Delta Y_{j=Greece, t-1}$ |
| Exogenous variable #2 | $\Delta Y_{j=Greece, t-2}$ | $\Delta Y_{j=Greece, t-2}$ | $\Delta Y_{j=Greece, t-2}$ |

Notes: the table reports Spruk and Kovac (2020) differential trend assumption test. Under the null hypothesis, the austerity policy does not induce a differential trend of the mortality difference between Greece and its synthetic control group, and both exhibit similar differenced mortality trend between pre- and post-austerity period. The table reports the p-values on the exact Chi-square test statistics of the Chow (1960) structural break in the triple-differences structural setup of the model. Two-sided p-values on the test statistics are reported in the parentheses.

Figure 10: Differential pre- and post-austerity trends in infant mortality gaps, 1991-2020

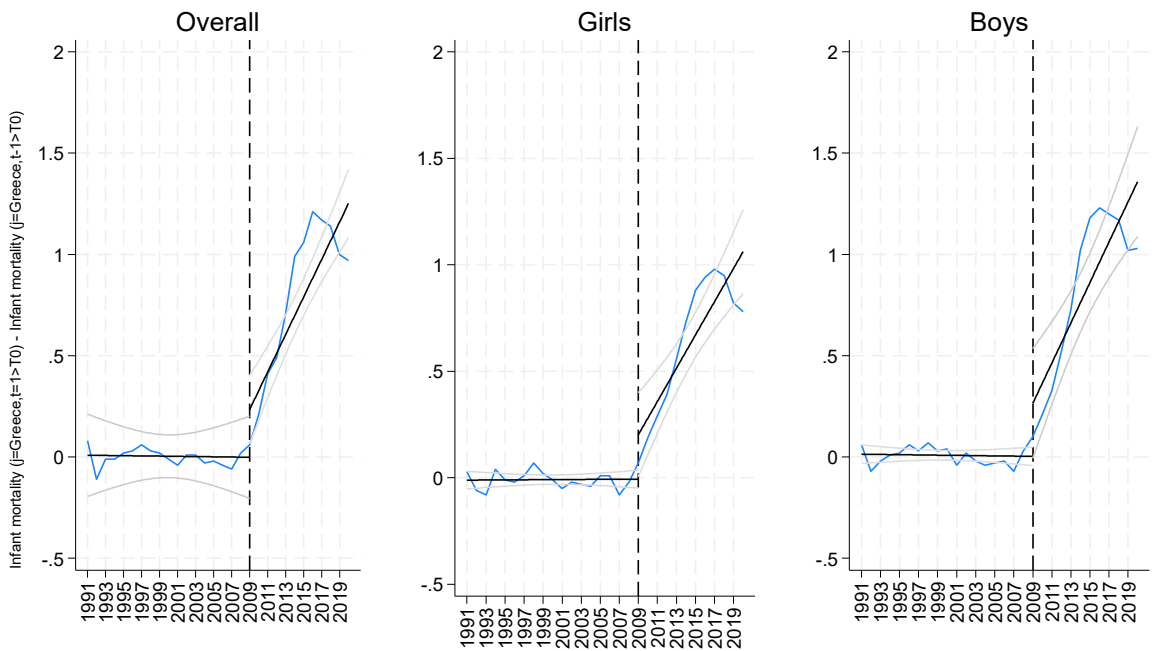

### 5.2.4 Leave-one-out analysis and sparsity conditions

Another internal validity concern of the estimated mortality effect is the composition of the synthetic control groups that best reproduce both the overall and gender-specific mortality trajectories prior t austerity. In the original and most parsimonious specification, the Greek mortality trajectory prior to 2009 is best reproduced by the weighted combination of convex attributes of Iceland, Estonia, Portugal and several others with minor weight shares. Among the donors with non-zero weight, Iceland represents almost one half of the synthetic control group. Around the same time as Greece, Iceland was severely hit by the economic and financial crisis with an estimated GDP decline by 7.6 percent in 2008 and 2.8 percent in 2009[†] which implies that SUTVA assumption of stable treatment value assignment may be partially violated. To address this particular concern, we re-estimate the treatment effect of austerity measures on infant mortality rates by performing leave-one-out analysis (Abadie, Diamond, and Hainmueller 2015; Klößner et al. 2018) and exclude Iceland as the major donor from the pool of comparison as it may posit potentially excessive and high-leverage influence on the magnitude and significance of the treatment effect of austerity measures.

Figure 11 presents leave-one-out re-evaluation of the treatment effect of austerity measures without Iceland in the donor pool. Overall, Greek infant mortality trajectory prior to the austerity policy is best reproduced through a convex combination of implicit attributes of Luxembourg (61 percent), Ireland (17 percent), Latvia (13 percent), and Estonia (9 percent), respectively. A slightly more nuanced but very similar composition of the synthetic control group is inherent in gender-disaggregated estimates of synthetic mortality curves. By and large, leave-one-out estimates of the mortality effect of austerity measures confirm our baseline estimates. More specifically, the average treatment effect of austerity policies on infant mortality is around 717 infant deaths in each year of austerity (p-value = 0.000). The cumulative infant mortality toll is around 7,891 infant deaths more than implied by the synthetic control group between 2010 and 2020 (i.e. end-of-sample p-value = 0.000). The average treatment effect of austerity on infant mortality of boys is around 27 percent higher than for girls. In a similar vein, the cumulative mortality toll for boys is 30 percent higher than for girls and the gender-specific difference is statistically significant at 1 percent (i.e. p-value = 0.000), and should be noted that the comparison of cumulative toll invariably implies that the boy death toll surpassed 10,000 while the girl death toll is around 7,891 in comparison with the synthetic control group. To tackle the aggregate uncertainty of the estimates, we invert the post-austerity test statistics through sparsity matrix of non-zero donors and compute the empirical 95% confidence intervals for each post-treatment year and all three mortality trajectories. Despite some uncertainty, it becomes apparent that post-austerity mortality gaps are both large and statistically significant, and confirm long-lasting mortality spike which appears somewhat stronger for boys. In spite of the shrinking mortality gap after 2015, large and statistically significant difference in mortality rates between Greece and its synthetic peers is prevalent up to the end-of-sample period.

---

[†] World Economic Outlook Database: April 2024
https://www.imf.org/en/Publications/WEO/weo-database/2024/April

**Figure 11**: Leave-one-out analysis of the infant mortality effect of austerity measures, 1990-2020

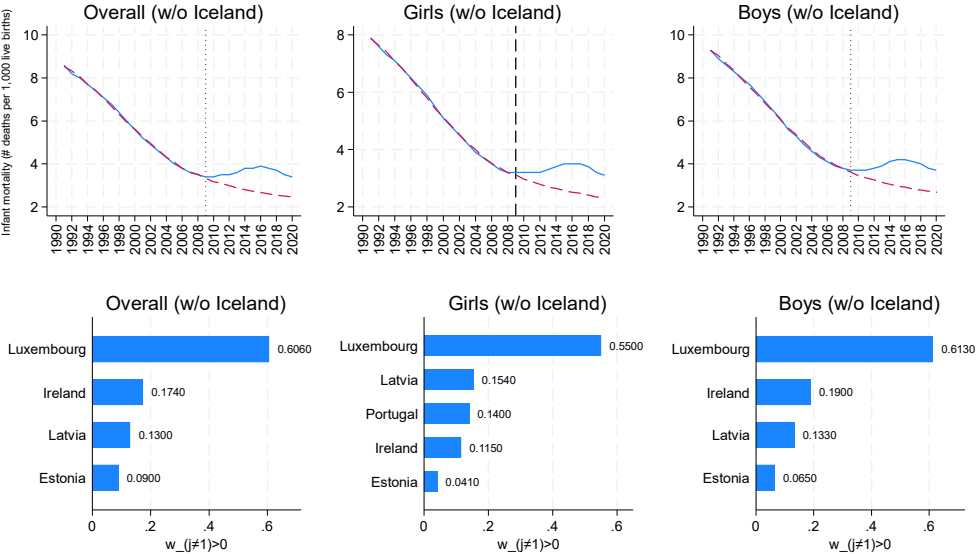

**Figure 12**: Effect of austerity measures on infant mortality under sparsity conditions, 1991-2020

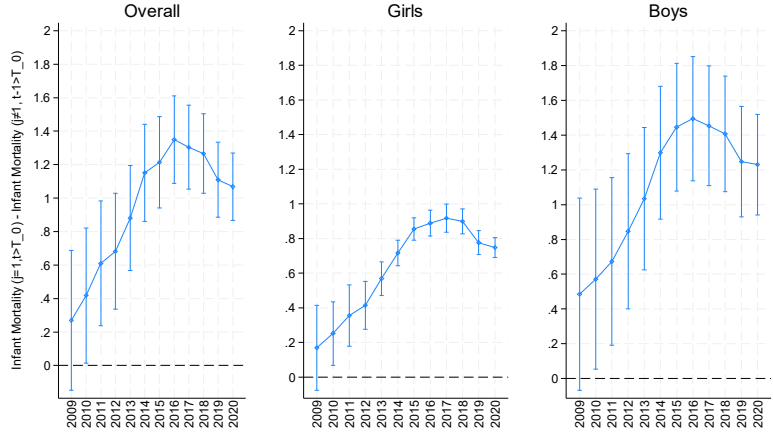

*5.2.4   Varying the composition of the donor pool*

A related but somewhat distinctive salient objection to the internal validity of our estimates may arise from the composition of the donor. A chief concern is posited by the comparison of Greece to an overly diverse donor pool of OECD countries where widely different trajectories and rates of infant mortality are perceptible in the range between 1.7 deaths per 1,000 live births in Estonia (in 2020) and 52 deaths per 1,000 live births in Türkiye in 1991. As pointed out by Abadie (2021), an overly diverse and large donor pool may increase the risk of over-fitting since a large size of the donor pool immediately translates into a larger number of discrepancies in factor loadings, biasing the synthetic control estimates. To expedite a judicious decision on the size of the donor pool based on the similarity of observed values of mortality within a common geographic space, we shrink the composition of the donor pool to the sample of Mediterranean states which belong to EUMED7 group within the European Union[‡] and Türkiye, which yields a substantially less diverse and more compact donor pool. Using a coarsened donor pool of Mediterranean states, we estimate the infant mortality effect of austerity measures using the classical synthetic control estimator and full-outcome path optimization.

Table 4 reports the overall and gender-disaggregated infant mortality effect of austerity measures using a reduced donor pool consisting of the Mediterranean states. The implicit attributes of the Mediterranean states in Europe provide a very good quality of fit with the pre-austerity infant mortality trajectories of Greece. In particular, the overall infant mortality trajectory of Greece in the pre-austerity period is best reproduced by the convex combination of the mortality rates of Cyprus (64 percent), Spain (30 percent), and Malta (6 percent), respectively. One half of the girls' mortality trajectory in the pre-austerity period can also be synthesized by Cyprus, followed by Italy (40 percent), Croatia (9 percent), and Malta (<1 percent). In a similar vein, the synthetic version of Greece in terms of boys' infant mortality dynamics prior to the austerity measures consists of Cyprus (43 percent), Slovenia (30 percent), Malta (10 percent), and Croatia (6 percent).

---

[‡] Croatia, Cyprus, France, Italy, Malta, Portugal, Slovenia and Spain

Table 4: Infant mortality effect of austerity measures in Greece using Mediterranean donor pool, 1990-2020

|  | Overall | Girls | Boys |
|---|---|---|---|
| Average mortality gap (p-value) | +0.857 | +0.733 | +0.878 |
|  | (0.000) | (0.000) | (0.000) |
| End-of-sample mortality gap (p-value) | +0.825 | +0.818 | +0.938 |
|  | (0.000) | (0.000) | (0.000) |
| RMSE | 0.229 | 0.172 | 0.291 |
| R2 | 0.98 | 0.98 | 0.97 |
| # control units | 10 | 10 | 10 |
| Bias | 0.15% | <0.1% | 0.15% |
| # pre-austerity outcomes | 19 | 19 | 19 |
| Composition of synthetic control groups |  |  |  |
| Croatia | 0 | 0.09 | 0.06 |
| Cyprus | 0.64 | 0.50 | 0.43 |
| France | 0 | 0 | 0 |
| Israel | 0 | 0 | 0 |
| Italy | 0 | 0.40 | 0 |
| Malta | 0.06 | <0.01 | 0.10 |
| Portugal | 0 | 0 | 0 |
| Slovenia | 0 | 0 | 0.30 |
| Spain | 0.30 | 0 | 0.11 |
| Türkiye | 0 | 0 | 0 |

The estimates show that the average mortality effect of the austerity measures in the post-intervention period is around +0.857 additional deaths per 1,000 live births (p-value = 0.000) whereas the boys mortality effect appears to be around 19 percent higher than the girls' mortality effect. By the end-of-sample period, the disparity in the gender-specific mortality effect lingers at 14 percent, and is statistically significant (p-value = 0.000) which confirms our prior findings. It should also be noted that the synthetic counterfactuals constructed from the Mediterranean donor pool of states yields the mortality gaps similar to our baseline of +0.85 basis point increase for the overall mortality, +0.69 basis point increase in girls' mortality, and +0.88 basis point increase in boys' mortality. Figure 13 reports the corresponding infant mortality gaps based on the Mediterranean donor pool of states.

**Figure 13**: Effect of austerity measures on infant mortality using the donor pool of European Mediterranean countries, 1991-2020

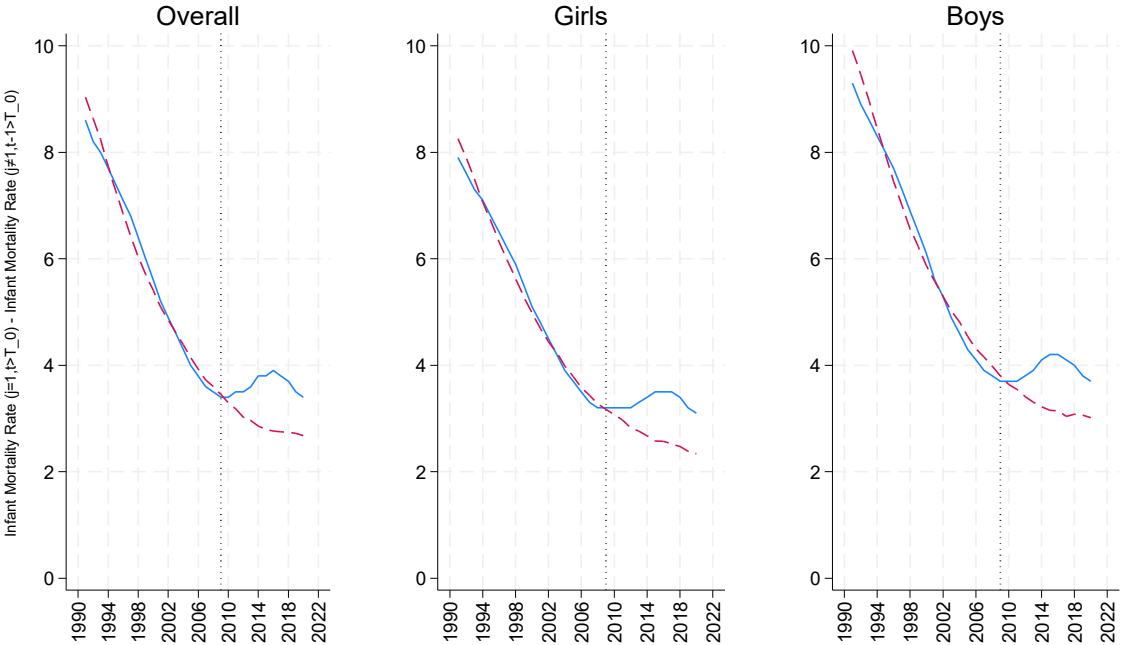

### 5.2.3 Generalized synthetic control estimates

The empirical analysis uncovers the evidence of substantial increase in infant mortality in response to the austerity policies. One of the main advantages of the synthetic control method lies in the easing of the parallel trend assumption between Greece and the donor pool of OECD countries given the uniqueness and severity of troika-imposed austerity measures. Yet, our model specification does not incorporate unobservable component into the projection and estimation of the counterfactual scenario whilst several challenges may arise when models with unobserved heterogeneity are estimated to evaluate the effect of austerity policy or related intervention of interest. To fill the void in the literature, Xu (2017) proposed a generalized version of the synthetic control method that unifies the counterfactual estimation based on the interaction of country-specific and year-specific effects to estimate the counterfactual (Xu 2017). Under this particular setup, the treatment effect is estimated using semi-parametric approach where the counterfactual for the treated unit is based on the linearly interactive fixed-effects model with time-varying coefficients interact with unit-specific intercepts. The respective counterfactual outcome trajectory is imputed from the observed outcome trajectory through a cross-validation that automatically selects the model with the best fit instead of arbitrary imposing covariates or pre-treatment outcomes. In turn, the risk of over-fitting is reduced the possibility of using the specification where many seemingly irrelevant auxiliary covariates are incorporated is mitigated inasmuch as possible. Compared to the classical synthetic control estimator, its generalized version uses the data from the pre-intervention period as a benchmark to derive the **reweighing scheme for the potential control units and yields the best possible prediction of the counterfactual scenario**.[§]

Table 5 reports interactive fixed-effects synthetic control estimates of the infant mortality effect of austerity measures for the period 1990-2020. More specifically, the table shows the average mortality differential relative to the synthetic counterfactual projected from the interactive of country- and time-specific effects in the data. The results confirm a large-scale and statistically significant increase in the mortality rate. Estimates reported in column (1) suggest that in the post-austerity period, overall infant mortality differential between Greece and its synthetic counterfactual is around 55 percent (p-value = 0.000). The estimated girls mortality differential is around 23 percent lower compared to boys whilst both differentials are statistically significant at 1% level. We tackle the relative uniqueness of the estimated mortality differentials by using the combined placebo test, performing both in-space permutation of the austerity policies to the unaffected countries as well as the assignment of the policy intervention to the deliberately wrong year. The evidence suggests that the increase in Greek infant mortality rates is unique and the placebo distribution of falsely assigned intervention does not appear to be similar to Greece, as we fail to reject the null hypothesis consistently. Furthermore, Figure 14 presents the estimated average treatment effect of austerity policy on infant mortality rate of Greece across the full post-intervention period. It also reports the in-time placebo analysis where the placebo year is selected through a cross-validation procedure indicating the year in which the post-intervention mortality increase may be most likely triggered by the year-specific events and policies. The evidence invariably suggests that the null hypothesis of in-time placebo effect cannot be rejected at conventional significance thresholds.

---

[§] In terms of further detail, the generalization of the classical synthetic control estimator proceeds in two distinctive steps. First, it allows for many treated units under the differential timing of the intervention. And second, the generalization also provides some estimate of uncertainty since standard errors and confidence bounds may be derived analytically and estimated empirically which eases the interpretation of the estimates considerably, further improving inference.

Table 5: Interactive fixed-effects synthetic control estimated infant mortality effect of austerity measures, 1990-2020

|  | Overall | Girls | Boys |
|---|---|---|---|
|  | (1) | (2) | (3) |
| A.T.T. | +1.552 | +1.338 | +1.725 |
|  | (.354) | (.343) | (.361) |
| Simulation-based p-value | 0.000 | 0.000 | 0.000 |
| Two-tailed 95% confidence interval | (1.269, 2.585) | (1.205, 2.325) | (1.384, 2.795) |
| Combined placebo test | YES | YES | YES |
| (p-value) | (0.723) | (0.702) | (0.472) |

Notes: the table reports the average treatment effect (ATT) of the austerity measures in Greece on infant mortality for the period 1990-2020 using interactive fixed-effects algorithm of the generalized synthetic control estimator. Standard errors are adjusted for serially correlated stochastic disturbances using cluster-robust error component model and finite empirical distribution function, and are reported in the parentheses.

Figure 14: Full placebo-region analysis of the infant mortality effect of austerity measures in Greece, 1990-2020

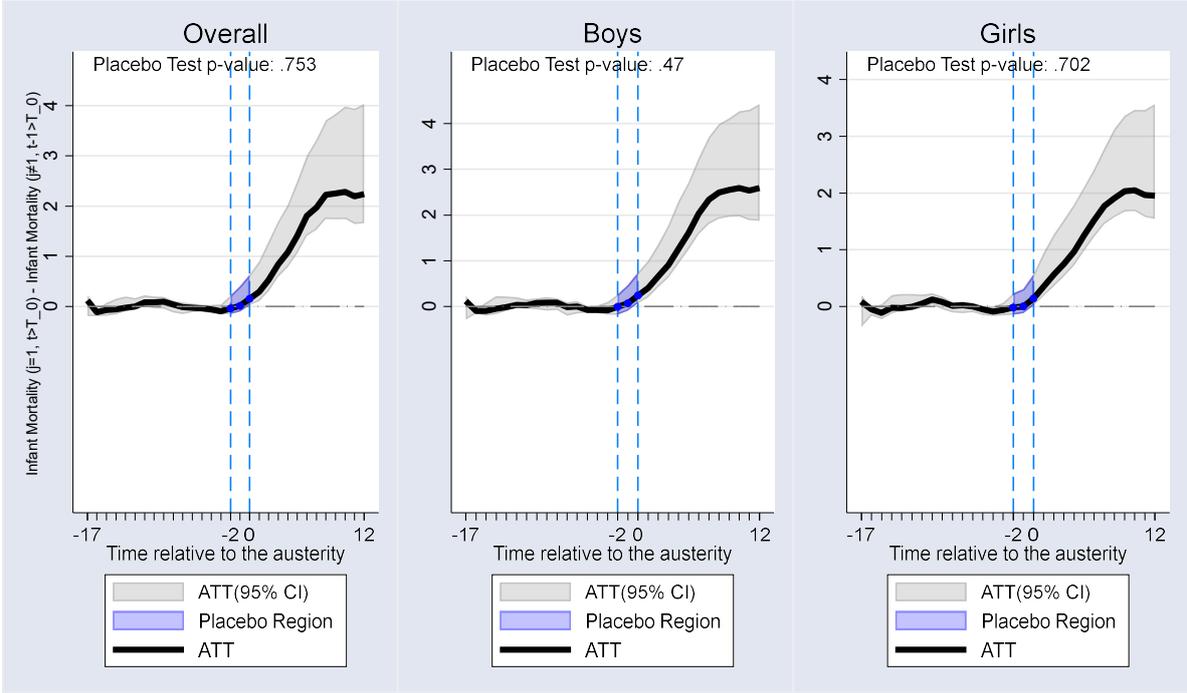

### 5.2.4 Synthetic difference-in-differences estimates

The comparison of mortality trajectories implies that Greece followed a parallel trend but yet experience a unique imposition of deep cuts in public health care spending. Thus, parallel trend assumption can be questioned. Instead of deliberately selecting the synthetic control estimator where parallel trend assumption is relaxed or difference-in-differences estimator where the assumption should hold, Arkhangelsky et. al. (2021) proposed synthetic difference-in-differences estimator leveraging a similar control group that approximates the outcome dynamics of the treatment group in the absence of the treatment (Arkhangelsky et al. 2021). The proposed estimator generates an optimally matched control group that loosens the necessity of parallel trend assumption, and relaxes both parallel trend assumption and the convex hull requirement to seek common support for the treated unit's characteristics based on the attributes within the convex hull. One of the notable advantages of the synthetic difference-in-differences estimator lies in its invariance to the additive shifts in the weight structure which poses a substantial improvement over synthetic control estimator. By balancing pre-intervention trends not only through country-level weights but also with time-varying weights, two-way fixed effects localized estimator can estimate the appropriate average treatment effect consistently.

Table 6 reports the magnitude of the overall and gender-specific infant mortality increase in Greece in response to the introduction of troika-imposed austerity policies in the period 2010-2020 based on the counterfactual imputed from country- and year-level weights. Specifically, the table reports the magnitude of the average treatment effect alongside two-tailed 95% confidence bounds and large-sample approximated empirical p-value on the null hypothesis. The evidence based on the synthetic difference-in-differences estimates confirms a sizeable increase in infant mortality in response to austerity measures by around 1.12 deaths per 1,000 live births relative to the estimated counterfactual (p-value = 0.003). Consistent with the earlier findings, the estimated effect is 14 percent higher for boys compared to girls. The evidence reaffirms the earlier results and indicates that the mortality increase in response to the austerity measures is both substantial and long-lasting. Figure 15 depicts synthetic difference-in-differences infant mortality effect along with the outcome diagnostic plot.

**Table 6**: Synthetic difference-in-differences estimated effect of austerity measures on infant mortality in Greece, 1991-2020

|  | Overall | Girls | Boys |
|---|---|---|---|
|  | (1) | (2) | (3) |
| $\gamma = IMR_{Greece, t>T_0} - IMR_{Synthetic, t>T_0}.$ | +1.128 | +1.032 | +1.186 |
|  | (0.375) | (0.338) | (0.669) |
| Large-sample approximated p-value | 0.003 | 0.002 | 0.076 |
| Two-tailed 95% confidence interval | (0.393, 1.863) | (0.368, 1.694) | (-0.125, 2.497) |

Notes: the table reports synthetic difference-in-differences estimated effect of the austerity measures on overall and gender-specific infant mortality rate (IMR) by matching Greece's pre-austerity infant mortality trajectories with a donor pool of 38 members of the Organization for Economic Cooperation and Development (OECD). The table reports the weighted difference between each designated Greek infant mortality trajectory and its control group based on the localized two-way fixed effect estimator using outcome model includes latent country-level factors interacted with latent time factors. It also reports large sample-approximated empirical p-value on the null hypothesis behind the average treatment effect. The lower and upper bound of the two-tailed 95% confidence interval and the standard error of the average treatment effect are reported in the parentheses.

**Figure 15**: Synthetic difference-in-differences adjusted infant mortality effect of austerity measures in Greece, 1991-2020

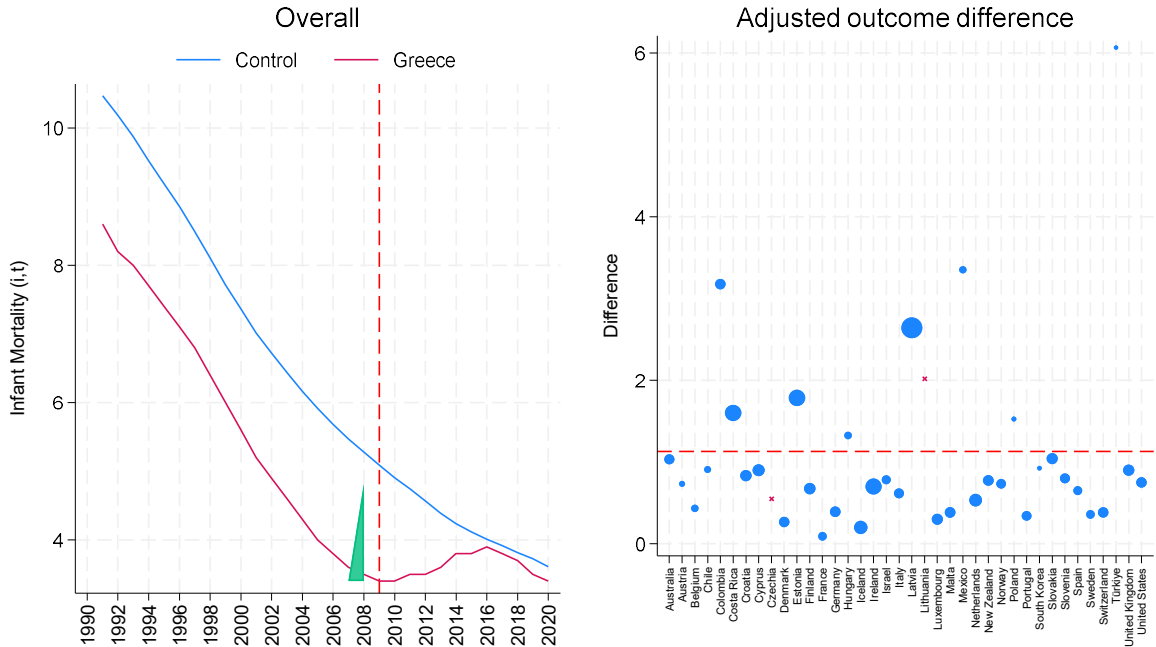

### 5.2.5 LASSO synthetic control estimates

Thus far, our empirical analysis rests on the intrinsic assumption of common support. This implies that the synthetic versions of Greece used to track, reproduce and synthesize Greek infant mortality trajectories consist of the weighted combination of OECD or Mediterranean countries attributes that belong to the convex hull of Greece's mortality trajectory based on the weights obtained in the training and validation stage. The common support assumption fundamentally allows for the extrapolation inside the convex hull of Greece and may be overly restrictive (Ben-Michael, Feller, and Rothstein 2021). To fill the void in the literature, Hollingsworth and Wing (2020) relax the convexity assumption and derive a more flexible version and regularized version of synthetic control estimator relying on LASSO-based machine learning algorithm that permits the extrapolation outside the convex hull (Hollingsworth and Wing 2020). This implies that both positive and negative weights for the treated unit's counterfactual can be estimated. The former indicate similarity with the treated unit whereas the latter designate some degree of dissimilarity. Such countercyclical weights without the additive restriction to unity facilitate improved flexibility and more nuanced estimation of the counterfactual with substantially reduced pre-treatment imbalance. If the exposure of treatment under potential outcomes satisfies conditional independence assumption and no major structural break is perceptible in the pre-intervention period, the treatment effect behind the austerity measures can be plausibly identified using a latent factor model.

Figure 16 shows the LASSO synthetic control estimated effect of austerity measures on overall and gender-specific infant mortality rate using two distinctive variants of the donor, namely, the coarsened donor pool of OECD countries and a more compact and salient donor pool of Mediterranean states from sub-section 5.2.4. A closer inspect of the pre-austerity mortality gaps self-suggests that it becomes apparent that the LASSO-supported synthetic control estimator provides an excellent fit of the mortality trajectories between Greece and its synthetic counterparts indicated both visually as well as judged on the basis on RMSE which is consistently below 1 percent of the pre-treatment error margin. The evidence confirms the overall infant mortality increase of around 1.16 additional deaths per 1,000 live births (p-value = 0.000). Under countercyclical weights, the estimated effect using the full OECD donor pool is comparable between boys and girls. The overall infant mortality dynamics of Greece prior to the austerity measures is best synthesized by the convex combination of attributes of Malta (26 percent), Ireland (21 percent), and Latvia (20 percent) alongside a handful of other OECD donors with smaller non-zero weight. By shrinking the donor pool to the Mediterranean states to capture more salient features of pre-austerity infant mortality of Greece, the synthetic control group loads positively on Malta (+42 percent), Cyprus (+25 percent), Türkiye (+15 percent), Portugal (+14 percent), Croatia (+2 percent), and Slovenia (<1 percent), and negatively on Italy (-10 percent),

France (-51 percent) and Spain (-52 percent). The estimated average increase in infant mortality implied from the comparison with other Mediterranean states is around 1.04 additional deaths per 1,000 live birth, respectively (i.e. p-value = 0.000). Aligned with our prior estimates, the magnitude of the average mortality gap is approximately 20 percent higher for boys compared to girls, and reiterates additional support for our baseline findings. Figure 17 reports the composition of overall and gender-disaggregated synthetic control groups under countercyclical weights in greater detail.

**Figure 16**: LASSO synthetic control estimated infant mortality effect of austerity measures in Greece, 1991-2020

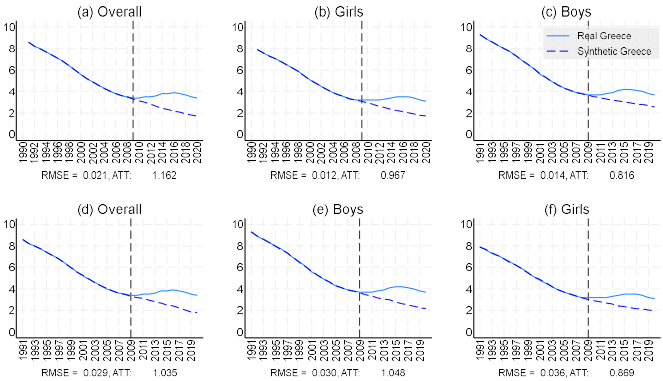

**Figure 17**: Composition of synthetic control groups across LASSO-based synthetic control estimation

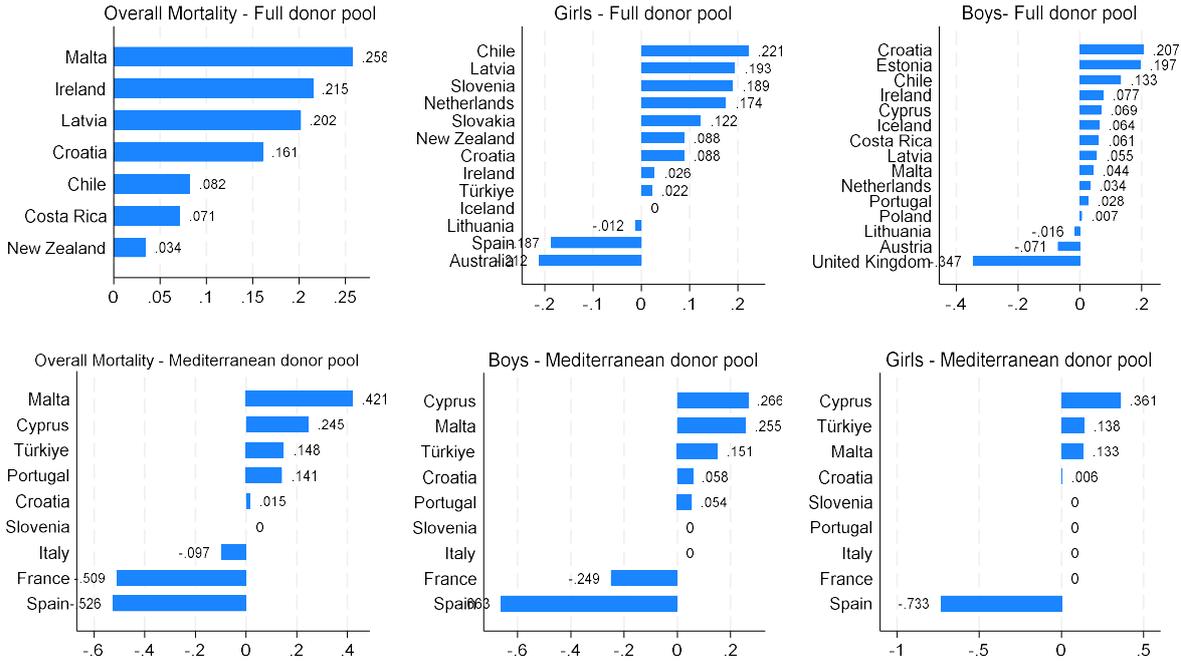

By way of example, the composition of the synthetic versions of Greece is similar to thatobtained in the baseline analysis of the infant mortality effect of austerity measures. In the overall mortality specification, the convexity requirement for ascribe common support for pre-austerity mortality dynamics is met. The synthetic version of Greece that best reproduces its pre-austerity mortality trajectory consists of Malta (26 percent), Ireland (22 percent), Latvia (20 percent), Croatia (16 percent), Chile (8 percent), Costa Rica (7 percent), and New Zealand (3 percent). Shrinking the OECD donor pool to the European sample of Mediterranean countries yields very similar composition of the synthetic version of Greece. In the overall mortality LASSO estimator with countercyclical weights, Greece's synthetic control group loads positively of Malta (42 percent), Cyprus (25 percent), Türkiye (15 percent), Portugal (14 percent), Croatia (2 percent) and negatively on Italy (-10 percent), France (-51 percent), and Spain (-53 percent) while Slovenia receives zero weight.

### 5.2.3 *Effects on neonatal and post-neonatal mortality*

One the remaining questions behind the estimated effect of austerity-imposed reduction in public health spending on infant mortality hinges on the broader generalization of the effect. To gauge the generalizability of the estimated effect, the outcome variable can be expanded to determine whether higher rates of mortality can be detected in response to austerity policies beyond the infancy stage. To address this question, we expand our analysis in two steps. First, we exploit the differences in neonatal mortality rate across OECD countries and use the synthetic control estimator to project the counterfactual scenario of neonatal mortality in the hypothetical absence of the austerity measures. Since the first month of life is considered to be the most vulnerable period for child survival and nearly one half of all deaths in children under the age of 5 occur in the first 28 days of life, neonatal mortality rate is an equally relevant indicator to considered in the evaluation of the effect of austerity policies on mortality rate. Premature birth, birth complications such as asphyxia and trauma as well as neonatal infections and congenital anomalies have been cited by World Health Organization as the leading causes of death in the neonatal stage. The question that arises immediately is whether austerity measures designed to curb the expenditure on public health care have affected neonatal mortality to the similar degree as infant mortality. And second, we also exploit cross-country differences in post-neonatal and under-five mortality rate and examine the contribution of austerity measures to the rate of mortality in the post-neonatal stage. By applying the synthetic control estimator to Greece and using the full OECD donor pool to ascribe the implicit attributes of mortality in the pre-austerity period, we estimate the respective counterfactual scenario.

Figure 18 illustrates the estimated neonatal mortality effect of the austerity measures. The point estimates of the mortality gap in the post-intervention period indicate a marked and permanently upward derailed mortality trajectory after the implementation of austerity measures. The estimated average increase in the mortality rate is around 0.798 (p-value) which is the equivalent of around 800 estimated live births ending with the death in the first 28 days of birth per each year of the austerity policies. The effect appears to be immediate and tends to increase gradually up until 2017 whilst remaining constant up to the end-of-sample post-treatment year. The synthetic version of Greece that best reproduces the neonatal mortality dynamics prior to the austerity package is consistent with our baseline composition, and consists of Iceland (48 percent), Estonia (33 percent), Ireland (13 percent), Latvia (5 percent), and Mexico (1 percent). The estimated increase in neonatal mortality is apparently permanent as the null hypothesis on the treatment effect of austerity policies can be very easily rejected for each post-intervention period. We also find evidence of elevated childhood mortality in the post-neonatal period up to the fifth year of life. By indicating the probability of death before the age of five under age-specific mortality rate, under-five mortality rate can be considered a baseline indicator of progress towards assuring children's rights to life, health care, nutrition, water, social security and protection. Our estimates indicate a severe deterioration of under-five mortality rate after the imposition of austerity measures. Pointwise, we find the average treatment effect of austerity around 0.788 (p-value = 0.000) which is very close to the baseline and neonatal estimate, and also is statistically significant at the conventional levels which indicates that around 92 percent of the deaths (=0.797/0.851) were concentrated in the infancy or neonatal stage, and the remainder in the post-neonatal stage. Furthermore, Figure 19 presents gender-disaggregated neonatal and post-neonatal mortality gaps, and shows that the in the post-neonatal stage, the average mortality increase in response to austerity is around 30 percent (=0.913/0.707) higher for boys compared to girls.

**Figure 18**: Post-infancy childhood mortality effect of austerity measures in Greece, 1991-2020

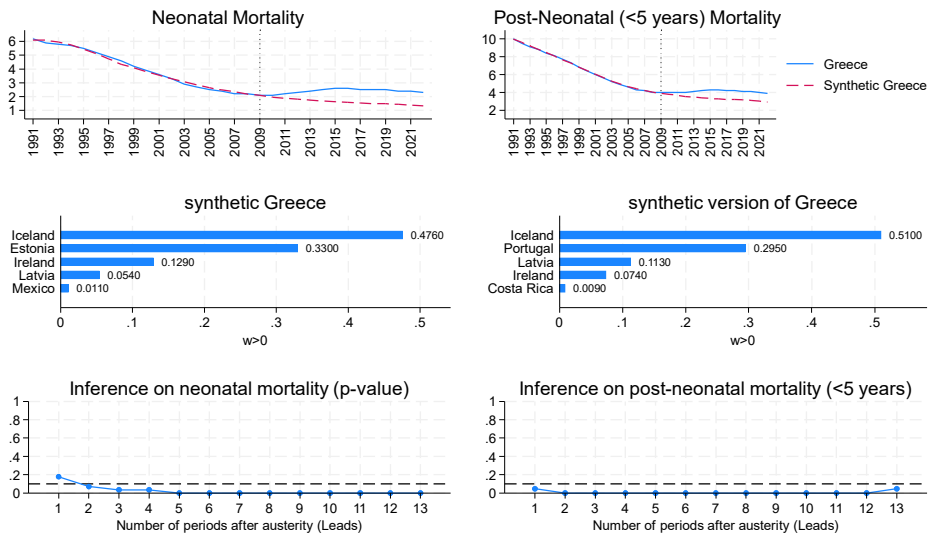

**Figure 19**: Gender-disaggregated effect of austerity policy on post-neonatal mortality in Greece, 1991-2020

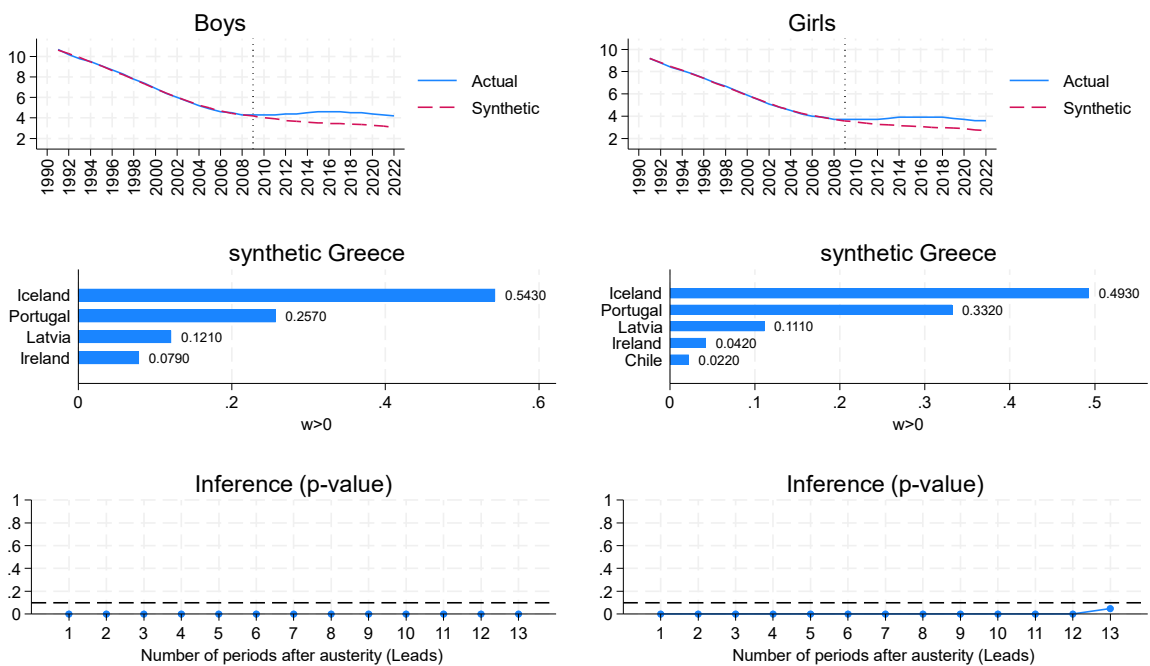

## 6 Conclusion

In this paper, we examine the contribution of austerity measures to the infant mortality. To this end, we exploit troika-led introduction of sharp cuts in public health spending in Greece in 2010 as a quasi-experimental source of variation in infant mortality and estimate the missing counterfactual scenario. Leveraging a relatively compact and optimally sized pre-austerity infant mortality trajectory of Greece against a stable donor pool of OECD countries where such cuts were never implemented, we reproduce the exact attributes of Greek infant mortality prior to the austerity, and estimate the missing counterfactual trajectory of infant mortality. Our evidence suggests that Greece's control group consisting of the implicit mortality attributes of the OECD countries prior to the austerity policies provides an excellent quality of the fit with no evidence of marked discrepancy. Our results suggest that the austerity policies resulted in the notable increase in infant mortality in response to deep and prolonged public health care spending reduction imposed by the austerity policies.

In particular, our synthetic control estimates indicate an increase in infant mortality rate around 43 percent in the post-austerity period. Relative to the estimated counterfactual scenario, the increase in infant mortality in response to the austerity measures does not appear and has persisted throughout the COVID19 pandemic with notable signs of a more permanent impact. The estimated impact of austerity measures appears to be around one third higher for boys compared to girls. The infant mortality gap survives an extensive battery of in-space and in-time placebo analysis as we show that by permuting the deep austerity measures either to the donor pool or to the deliberately wrong policy years, the increase in infant mortality appears to be unique and persists up to the present day. A closer inspection of the mortality gaps indicates that the austerity policies triggered a differential trend in the trajectory of infant mortality that appears to be statistically significant at conventional levels. Furthermore, the increase in infant mortality is robust against the varying composition of the donor pool as well as to the extrapolation outside the convex hull using countercyclical negative weights. By examining a wider range of child mortality outcomes, we should that around 92 percent of the infant deaths were concentrated in the infancy and neonatal period whilst the remaining deaths unfolded in the post-neonatal period.

Compared to earlier studies, our approach and results precisely quantify the mortality toll relative to the counterfactual under which sharp cuts in public health expenditure are not implemented. It is almost unbelievable but unfortunately true to see the earlier results reporting an increase in HIV and tuberculosis incidence after 2009 although the latter two were relatively short-lived. However, infant mortality trajectory appears to have been permanently derailed upward up until COVID19 pandemic. Against the backdrop of existing findings in the literature, our results show that extensive austerity in the economic downturn impedes healthcare quality and service access, particularly among vulnerable and fragile populations such as disabled children and suggest that the social consequences of austerity measures will be long-lasting.